\documentclass[prd,aps,notitlepage,showpacs,nofootinbib,tightenlines]{revtex4}
\usepackage{mathrsfs}
\usepackage{amsmath}
\usepackage{amssymb}
\usepackage{epsfig}
\usepackage{graphicx}
\usepackage{booktabs}
\usepackage{multirow}
\usepackage{subfigure}
\usepackage{bm}
\usepackage{times}
\usepackage{braket}
\usepackage{color}
\usepackage{slashed}
\usepackage{hyperref}
\usepackage{threeparttable}
\newcommand{\beq}{\begin{eqnarray}}
\newcommand{\eeq}{\end{eqnarray}}
\newcommand{\non}{\nonumber\\ }

\definecolor{Red}{rgb}{1.,0.,0.}

\definecolor{Blue}{rgb}{0.,0.,1.}
\newcommand{\Blue}[1]{{\color{Blue}{#1}}}

\definecolor{nicered}{rgb}{0.7,0.1,0.1}
\definecolor{nicegreen}{rgb}{0.1,0.5,0.1}
\hypersetup{colorlinks,citecolor=nicegreen,linkcolor=nicered}
\def \cpc{ Chin. Phys. C  }

\def \epjc{ Eur. Phys. J. C }

\def \jpg{  J. Phys. G }
\def \npb{  Nucl. Phys. B }
\def \plb{  Phys. Lett. B }

\def \prd{  Phys. Rev. D }
\def \prl{  Phys. Rev. Lett.  }

\def \jhep{ J. High Energy Phys. }
\def \B{{\cal B}}

\begin{document}

\title{Study of $B_{(s)} \to (\pi\pi)(K\pi)$ decays in the perturbative QCD approach}

\author{Ya Li$^1$}                \email[]{liyakelly@163.com (corresponding author)}
\author{Da-Cheng Yan$^2$}         \email[]{yandac@126.com (corresponding author)}
\author{Zhou Rui$^3$}              \email[]{jindui1127@126.com}
\author{Zhen-Jun Xiao$^{4}$}    \email[]{xiaozhenjun@njnu.edu.cn}
\affiliation{$^1$ Department of Physics, College of Sciences, Nanjing Agricultural University, Nanjing, Jiangsu 210095, China}
\affiliation{$^2$ School of Mathematics and Physics, Changzhou University, Changzhou, Jiangsu 213164, China}
\affiliation{$^3$ College of Sciences, North China University of Science and Technology, Tangshan, Hebei 063210, China}
\affiliation{$^4$ Department of Physics and Institute of Theoretical Physics,
                          Nanjing Normal University, Nanjing, Jiangsu 210023, China}
\date{\today}

\begin{abstract}
In this work, we provide estimates of the branching ratios, direct $CP$ asymmetries and triple product asymmetries in $B_{(s)} \to (\pi\pi)(K\pi)$ decays in the perturbative QCD approach, where the $\pi\pi$ and $K\pi$ invariant mass spectra are dominated by the vector resonances $\rho(770)$ and $K^*(892)$, respectively.
Some scalar backgrounds, such as $f_0(500,980) \to \pi\pi$ and $K^*_0(1430) \to K\pi$ are also accounted for.
The $\rho(700)$ is parametrized by the Gounaris-Sakurai function.
The relativistic Breit-Wigner formula for the $f_0(500)$ and Flatt\'e model for the $f_0(980)$ are adopted to parameterize the time-like scalar form factors $F_S(\omega^2)$.
We also use the D.V.~Bugg model to parameterize the $f_0(500)$ and compare the relevant theoretical predictions from
different models.
While in the region of $K\pi$ invariant mass, the $K^*_0(1430)$ is described with the LASS lineshape and the $K^*(892)$ is modeled by the Breit-Wigner function.
We find that the decay rates for the considered decay modes agree with currently available data within errors.
As a by-product, we extract the branching ratios of two-body decays $B_{(s)} \to \rho(770)K^*(892)$ from the corresponding four-body decay modes and calculate the relevant polarization fractions.
Our prediction of longitudinal polarization fraction for $B^0\to \rho(770)^0 K^*(892)^0$ decay deviates a lot from the recent LHCb measurement, which should be resolved.
It is shown that the direct $CP$ asymmetries are large due to the sizable interference between the tree and penguin contributions, but they are small for the tree-dominant or penguin-dominant processes.
The PQCD predictions for the ``true" triple product asymmetries are small which are expected in the standard model, and consistent with the current data reported by the LHCb Collaboration.
Our results can be tested by the future precise data from the LHCb and Belle II experiments.
\end{abstract}

\maketitle

\section{Introduction}
\begin{figure}[tbp]
\centerline{\epsfxsize=9cm \epsffile{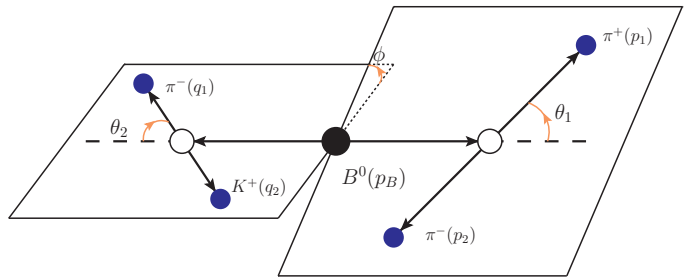}}
\caption{Graphical definitions of the helicity angles $\theta_1$, $\theta_2$ and $\phi$.
Taking the $B^0 \to R_1R_2$ decay as an example, with each
quasi-two-body intermediate resonance decaying to two pseudoscalars ($R_1\to \pi^+\pi^-$ and $R_2\to K^+\pi^-$), $\theta_1(\theta_2)$ is denoted as the angle between the directions of motion of $\pi^+$ ($\pi^-$) in the $R_1(R_2)$ rest frame and $R_1(R_2)$ in the $B^0$ rest frame, and $\phi$ is the angle between the plane defined by $\pi^+\pi^-$ and the plane defined by $K^+\pi^-$ in the $B^0$ rest frame.}
\label{fig1}
\end{figure}

Multi-body $B$ meson decays offer one of the best tools for studying direct $CP$ violation and provide an interesting testing ground for strong interaction dynamical models.
Theoretically, $B_{(s)} \to V_1V_2, S_1S_2, S_1V_2, V_1S_2$ decays (here $V_{1,2}$ and $S_{1,2}$ denote the vector and scalar mesons, respectively) are treated as two-body final states and have been studied in the two-body framework using QCD factorization (QCDF)~\cite{npb774-64,prd77-014034,prd80-114008,prd80-114026,1206-4106,prd87-114001}, the perturbative QCD (PQCD) approaches~\cite{plb622-63,prd72-054015,jpg32-101,prd73-014011,prd74-114010,prd73-104024,prd76-074018,prd81-074014,epjc67-163,prd82-034036,prd88-094003,prd91-054033,npb935-17}, the soft-collinear-effective theory (SCET)~\cite{prd70-054015,prd74-034010,npb692-232,prd72-098501,prd72-098502,prd96-073004} and the factorization-assisted topological amplitude approach (FAT)~\cite{epjc77-333}.
While they are at least four-body decays on the experimental side shown in Fig.~\ref{fig1}, since the vector (scalar) mesons decay via the strong interaction with a nontrivial width.
In recent years, four-body charmless hadronic $B$ decays have been reported by BABAR~\cite{prl97-201801,prd78-092008,prl100-081801,prl101-201801,prd83-051101,prd85-072005}, Belle~\cite{prl95-141801,prd80-051103,prd81-071101,prd88-072004} and LHCb~\cite{plb709-50,jhep11-092,jhep05-069,jhep07-166,jhep10-053,jhep03-140,jhep05-026,jhep07-032} Collaborations, and the branching ratios for many partial waves have been measured, see Tables~\ref{exp1} and \ref{exp2} for a summary of the experimental results.
In addition to the branching ratios and the polarization fractions, other observables constructed from the helicity amplitudes are also interesting.
The phenomenology of these decay modes provides rich opportunities for our understanding of the mechanism for hadronic weak decays and their $CP$ asymmetry, and the search for physics beyond the standard model (SM).

Besides the direct $CP$ asymmetries, there is another signal of $CP$ violation in the angular distribution of $B_{(s)} \to V_1V_2$ decays, which is called triple-product asymmetries (TPAs)~\cite{prd39-3339,npbps13-487,ijmpa19-2505,prd84-096013,plb701-357,prd86-076011,prd88-016007,prd92-076013,prd87-116005}.
These triple products are odd under the time reversal transformation ($T$), and also contribute potential signals of $CP$ violation due to the $CPT$ theorem.
TPAs have already been measured by BABAR, Belle, CDF and LHCb~\cite{prd76-031102,prl95-091601,prl107-261802,plb713-369,prd90-052011,jhep07-166,jhep05-026}.
It is known that a non-vanishing direct $CP$ violation needs the interference of at least two amplitudes with a weak phase difference $\Delta \phi$ and a strong phase difference $\Delta \delta$.
The direct $CP$ violation is proportional to $\sin\Delta \phi \sin \Delta \delta$, while TPAs go as $\sin\Delta \phi \cos \Delta \delta$.
If the strong phases are quite small, the magnitude of the direct $CP$ violation is close to zero, but the TPA is maximal.
Hence direct $CP$ violation and TPAs complement each other.
Even if the effect of $CP$ violation is absent, $T$-odd triple products (also called ``fake" TPAs), which are proportional to $\cos\Delta \phi \sin \Delta \delta$, can provide useful complementary information on new physics~\cite{plb701-357}.
TPAs are excellent probes of physics beyond the SM since most TPAs are expected to be tiny within the SM and are not suppressed by the small strong phases.

As is well known, the kinematics of two-body decays is fixed, while the multi-body decay amplitudes depend on at least two kinematic variables.
Meanwhile, the multi-body decays not only receive the resonant and nonresonant contributions, but also involve the
possible significant final-state interactions (FSIs)~\cite{prd89-094013,1512-09284,prd89-053015}.
In this respect, multi-body decays are considerably more challenging than two-body decays, but provide a number of theoretical and phenomenological advantages.
In two-body $B$ decays, the measured $CP$ violation is just a number while the $CP$ asymmetry depends on the invariant mass
of the two-meson pair and varies from region to region in the Dalitz plot~\cite{dalitz-plot1,dalitz-plot2} in the three-body modes~\cite{prd90-112004}.
In addition, strong phases in multi-body decays arise nonperturbatively already at the leading power, through complex phases in matrix elements such as $F_{\pi}\sim <0|j|\pi\pi>$ and so on.
Since the $B_{(s)} \to V_1V_2, V_1S_2, S_1V_2, S_1S_2$ decays are expected to proceed through $V_1(S_1) \to P_1P_2$ and $V_2(S_2) \to Q_1Q_2$ with $P_{1,2}$ ($Q_{1,2}$) denoting pseudoscalar meson $\pi$ or $K$, it is meaningful to study such decays in the four-body framework, which provide useful information for understanding the $CP$-violation mechanisms.

As addressed above, multi-body decays of heavy mesons involve more complicated dynamics than two-body decays.
A factorization formalism that describes a multi-body decay in full phase space is not yet available at present.
It has been proposed that the factorization theorem of three-body $B$ decays is approximately valid when two particles move collinearly and the bachelor particle recoils back~\cite{plb561-258,prd79-094005}.
More details can also be found in Refs.~\cite{1609-07430,npb899-247}.
This situation exists particularly in the low $\pi\pi$ or $K\pi$ invariant mass region ($\lesssim$2 GeV) of the Dalitz plot where most resonant structures are seen.
The Dalitz plot is typically dominated by resonant quasi-two-body contributions along the edge.
This proposal provides a theoretical framework for studies of resonant contributions based on the quasi-two-body-decay mechanism.
Several theoretical approaches have been developed for describing the three-body  hadronic decays of $B$ mesons based on the symmetry principles~\cite{prd72-094031,plb727-136,prd72-075013,prd84-056002,plb728-579,prd91-014029}, the QCDF~\cite{plb622-207,prd74-114009,APPB42-2013,prd76-094006,prd88-114014,prd94-094015,prd89-094007,prd87-076007,jhep10-117,2005-06080,prd99-076010} and the PQCD approaches~\cite{plb763-29,prd95-056008,prd96-093011,prd98-056019,prd98-113003,jpg46-095001,cpc43-073103,epjc79-37,cpc44-073102,jhep03-162,epjc80-394,2005-02097,prd101-016015,prd102-056017,2105-03899}.
Recently, the localized $CP$ violation and branching fraction of the four-body decay
$\bar{B}^0\to K^-\pi^+\pi^+\pi^-$ have been calculated by employing a quasi-two-body QCDF
approach in Refs.~\cite{1912-11874,2008-08458}.
Similar to three-body $B$ meson decays, four-body $B_{(s)} \to R_1R_2\to (P_1P_2)(Q_1Q_2)$
decay modes ($R_{1,2}$ represents the vector or scalar intermediate resonance) are assumed to proceed dominantly with two intermediate resonances $R_1$ or $R_2$ each decaying to a pseudoscalar pair.
As a first step, we can only restrict ourselves to the specific kinematical configurations in which each two particles move collinearly and two pairs of final state particles recoil back in the rest frame of the $B$ meson, see Fig.~\ref{fig1}.
Naturally the dynamics associated with the pair of final state mesons can be factorized into a two-meson distribution amplitude (DA)
$\Phi_{h_1h_2}$~\cite{MP,MT01,MT02,MT03,NPB555-231,Grozin01,Grozin02}.
Thereby, the typical PQCD factorization formula for the considered four-body decay amplitude can be described as the form of,
\begin{eqnarray}
\mathcal{A}=\Phi_B\otimes H\otimes \Phi_{P_1P_2}\otimes\Phi_{Q_1Q_2},
\end{eqnarray}
where $\Phi_B$ is the universal wave function of the $B$ meson and absorbs the non-perturbative dynamics in the process.
The $\Phi_{P_1P_2,Q_1Q_2}$ is the two-hadron DA, which involves the resonant and nonresonant interactions between the two moving collinearly mesons.
The hard kernel $H$ describes the dynamics of the strong and electroweak interactions in four-body hadronic decays in a similar way as the one for the corresponding two-body decays.

\begin{figure}[tbp]
\centerline{\epsfxsize=14cm \epsffile{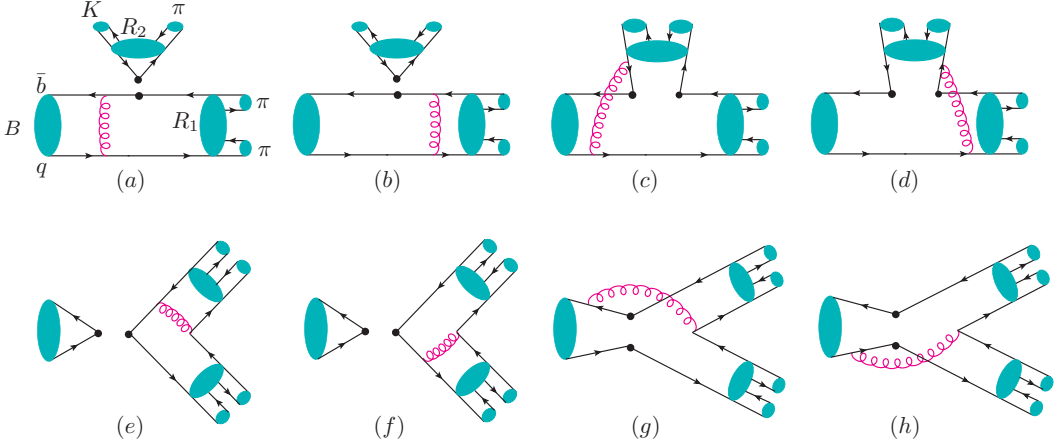}}
\caption{Typical leading-order Feynman diagrams for the four-body decays $B \to (R_1\to) \pi\pi (R_2\to) K\pi$ with $q=(u,d,s)$, where the symbol $\bullet$ denotes a weak interaction vertex. The diagrams ($a$)-($d$) represent the $B\to (R_1\to) \pi\pi$ transition, as well as the diagrams ($e$)-($h$) for annihilation contributions. If we exchange the position of $R_1(\to \pi\pi)$ and $R_2 (\to K\pi)$, we will find the diagrams ($a$)-($d$) for the $B\to (R_2\to) K\pi$ transition.}
\label{fig2}
\end{figure}

In this work, we study the four-body decays $B_{(s)} \to (\pi\pi)(K\pi)$ in the PQCD approach based on $k_T$ factorization with the relevant Feynman diagrams illustrated
in Fig.~\ref{fig2}.
In the considered $(\pi\pi)$ invariant-mass range, the vector resonance $\rho(770)$ is expected to contribute, together with the scalar resonances $f_0(500)$ and $f_0(980)$.
The $K\pi$ spectrum is dominated by the vector $K^*(892)$ resonance and the scalar resonance $K_0^*(1430)$.
Throughout the remainder of the paper, the symbol $\rho$ is used to denote the $\rho(770)$ resonance and $K^*$ is for $K^*(892)$ resonance.
For a comparison with the LHCb experiment~\cite{jhep05-026}, the invariant mass of the $\pi\pi$ pair is restricted from 300 MeV to 1100 MeV and the range of invariant mass for $K\pi$ pair varies from 750 MeV to 1200 MeV.
We calculate the branching ratios and polarization fractions of each partial waves.
Besides, a set of $CP$-violating observables is investigated using $B_{(s)}$ meson decays reconstructed in the $(\pi\pi)(K\pi)$ quasi-two-body final state.
Particular emphasis is placed on the TPAs of the considered decays.
It should be mentioned that there is a possibility of existing two identical final state pions.
Taking the $B^+ \to (f_0(980) \to )\pi^+\pi^-(K^{*+}\to)K^0\pi^+$ decay for an example, in the experimental side, they see four final states $\pi^+_{1}\pi^-K^0\pi^+_{2}$ first, and have to pair one of the positive charged
pions  with the $\pi^-$. They have to try two possible combinations.
From the theoretical point of view, we deal with $B^+ \to f_0(980) K^{*+}\to \pi^+_{1}\pi^-K^0\pi^+_{2}$ in the quasi-two-body mechanism and see $f_0(980)K^{*+}$ first.
Then each quasi-two-body intermediate resonance decays into two pseudoscalars ($f_0(980)\to \pi^+_{1}\pi^-$ and $K^{*0}\to K^0\pi^+_{2}$), respectively.
Each meson pair generates a smaller invariant mass and flies  back-to-back as shown in Fig.~\ref{fig1}.
It is evident that the final states $\pi^+$ have been already specified unambiguously
on the theoretical side. Certainly, it is experimentally difficult
to identify which resonance a $\pi^+$ comes from in the nonresonant
region on a Dalitz plot. From the theoretical viewpoint, this
difficulty implies that the tail of our Breit-Wigner propagator
may not describe the events in the nonresonant region correctly.
However, it is known that the nonresonant region gives a minor
contribution, so the misparing of a $\pi^+$ with other final states
should not make a significant impact to our results.

The layout of the present paper is as follows.
In Sec.~\ref{sec:2}, we give an introduction for the theoretical framework.
The numerical values and some discussions will be given in Sec.~\ref{sec:3}.
Section~\ref{sec:4} contains our conclusions.
The Appendix collects the explicit PQCD factorization formulas for all the decay amplitudes.

\begin{table}[!]
\caption{Experimental branching fractions (in units of
$10^{-6}$) of $B \to (\pi\pi)(K\pi)$ decays~\cite{1909-12524,pdg2020}.
} \label{exp1}
\begin{center}
\begin{threeparttable}
\begin{tabular}{l c c c c}
\hline \hline
Mode & BABAR & Belle & HFLAV &PDG2020 \\
 \hline
 $\B(B^+\to \rho^0 K^{*}(892)^+)$&$4.6\pm1.0\pm0.4$~~\cite{prd83-051101}&&$4.6\pm1.1$ &$4.6\pm1.1$\\
  $\B(B^+\to \rho^+K^{*}(892)^0)$&$9.6\pm1.7\pm1.5$~~\cite{prl97-201801}&$8.9\pm1.7\pm1.2$~~\cite{prl95-141801}&$9.2\pm1.5$&$9.2\pm1.5$\\
  $\B(B^+\to f_0(980)K^{*}(892)^+)\times\B(f_0(980)\to\pi^+\pi^-)$ &$4.2\pm0.6\pm0.3$~~\cite{prd83-051101} &&$4.2\pm0.7$  &  $4.2\pm0.7$ \\
  $\B(B^0\to \rho^0 K^{*}(892)^0)$&$5.1\pm0.6^{+0.6}_{-0.8}$~~\cite{prd85-072005}&$2.1^{+0.8+0.9}_{-0.7-0.5}$~~\cite{prd80-051103}&$3.9\pm0.8$&$3.9\pm1.3$\\
  $\B(B^0\to \rho^-K^{*}(892)^+)$&$10.3\pm2.3\pm1.3$~~\cite{prd85-072005}&&$10.3\pm2.6$&$10.3\pm2.6$\\
    $\B(B^0\to \rho^0 K_0^{*}(1430)^0)$&$27\pm4\pm2\pm3$~~\cite{prd85-072005}&&$27\pm5$&$27\pm6$\\
  $\B(B^0\to \rho^- K_0^{*}(1430)^+)$&$28\pm10\pm5\pm3$~~\cite{prd85-072005}&&$28\pm11$&$28\pm12$\\
   $\B(B^0\to f_0(980)K^*(892)^0)\times\B(f_0(980)\to\pi\pi)$ &$5.7\pm0.6\pm0.4$~~\cite{prd85-072005}&$1.4^{+0.6+0.6}_{-0.5-0.4}$~~\cite{prd80-051103}&$3.9\pm0.5$&$3.9^{+2.1}_{-1.8}$\\
   $\B(B^0\to f_0(980) K^{*}_0(1430)^0)\times\B(f_0(980)\to\pi\pi)$&$2.7\pm0.7\pm0.5\pm0.3$~~\cite{prd85-072005}&&$2.7\pm0.9$&$2.7\pm0.9$\\
     $\B(B_s^0\to \rho^0 \bar{K}^*(892)^0)$&&&&$<767$\\
 \hline \hline
\end{tabular}
\end{threeparttable}
\end{center}
\end{table}

\begin{table}[!]
\caption{Experimental $CP$ asymmetries (in units of \%) of $B \to (\pi\pi)(K\pi)$ decays~\cite{1909-12524,pdg2020}.
} \label{exp2}
\begin{center}
\begin{tabular*}{13cm}{@{\extracolsep{\fill}}l c c c c}
\hline \hline
Mode & BABAR  & HFLAV &PDG2020 \\
 \hline
 $B^+\to \rho^0 K^{*}(892)^+$&$31\pm13\pm3$~~\cite{prd83-051101}&$31\pm13$ &$31\pm13$\\
  $B^+\to \rho^+K^{*}(892)^0$&$-1\pm16\pm2$~~\cite{prl97-201801}&$-1\pm16$&$-1\pm16$\\
  $B^+\to f_0(980)K^{*}(892)^+$ &$-15\pm12\pm3$~~\cite{prd83-051101} &$-15\pm12$  &  $-15\pm12$ \\
  $B^0\to \rho^0 K^{*}(892)^0$&$-6\pm9\pm2$~~\cite{prd85-072005}&$-6\pm9$&$-6\pm9$\\
  $B^0\to \rho^-K^{*}(892)^+$&$21\pm15\pm2$~~\cite{prd85-072005}&$21\pm15$&$21\pm15$\\
   $B^0\to f_0(980)K^{*0}(892)$ &$7\pm10\pm2$~~\cite{prd85-072005}&$7\pm10$&$7\pm10$\\
 \hline \hline
\end{tabular*}
\end{center}
\end{table}

\section{FRAMEWORK}\label{sec:2}
\subsection{Kinematics}
\begin{table}  
\caption{Parameters used to describe intermediate states in our framework. GS and RBW refer to the Gounaris-Sakurai and relativistic Breit-Wigner line shapes, respectively.}
\label{Tab:pa}
\begin{center}
\begin{tabular*}{12cm}{@{\extracolsep{\fill}}llllll} \hline\hline
{\rm Resonance} &Mass~[MeV] &Width~[MeV] &$J^{P}$&Model&Source\\ \hline
$f_0(500)$ &$471\pm21$ &$534\pm53$ &$0^+$&\rm RBW  &LHCb~\cite{prd89-092006}\\
$f_0(980)$ &$965\pm8$ &$-$ &$0^+$&\rm Flatt\'e &BES~\cite{plb607-243}\\
$\rho(770)$ &$775.02\pm0.31$ &$149.59\pm0.67$ &$1^-$&\rm GS &BABAR~\cite{prd86-032013}\\
$K^*(892)$      &$895.55\pm0.20$&$47.3\pm0.5$&$1^{-}$&\rm RBW&LHCb~\cite{epjc78-1019}\\
$K_0^*(1430)$    &$1425\pm 50$&$270\pm80$&$0^{+}$&\rm LASS&LHCb~\cite{epjc78-1019}\\
\hline\hline
\end{tabular*}
\end{center}
\end{table}

Consider a four-body $B(p_B)\to R_1(p)R_2(q)\to P_1(p_1)P_2(p_2)Q_1(q_1)Q_2(q_2)$ decay, we will work in the $B$ meson rest frame and employ the light-cone coordinates for momentum variables.
The $B$ meson momentum $p_{B}$, the total momenta of the pion-pion and kaon-pion pairs,
$p=p_1+p_2$,  $q=q_1+q_2$, and the quark momentum $k_i$ in each meson are chosen as
\begin{eqnarray}
p_{B}&=&\frac{m_{B}}{\sqrt 2}(1,1,0_{\rm T}),~\quad k_{B}=\left(0,x_B \frac{m_{B}}{\sqrt2} ,k_{B \rm T}\right),\non
p&=&\frac{m_{B}}{\sqrt2}(g^+,g^-,0_{\rm T}),~\quad k_p= \left( x_1 g^+\frac{m_{B}}{\sqrt2},0,k_{\rm T}\right),\non
q&=&\frac{m_{B}}{\sqrt 2}(f^-,f^+,0_{\rm T}), ~\quad k_q=\left(0,x_2f^+ \frac{m_B}{\sqrt{2}},k_{3{\rm T}}\right),\label{mom-B-k}
\end{eqnarray}
where $m_{B}$ is the mass of $B$ meson. The momentum fractions $x_{B}$, $x_1$ and $x_2$ run from zero to unity.

The factors $g^{\pm}$ and $f^{\pm}$ can be obtained through the momentum conservation $p_B=p+q$, and $p^2=\omega_1^2$ and $q^2=\omega_2^2$, respectively.
The explicit expressions of $g^{\pm}$ and $f^{\pm}$ are written in the following form
 \begin{eqnarray}\label{eq:epsilon}
g^{\pm}&=&\frac{1}{2}\left[1+\eta_1-\eta_2\pm\sqrt{(1+\eta_1-\eta_2)^2-4\eta_1}\right],\nonumber\\
f^{\pm}&=&\frac{1}{2}\left[1-\eta_1+\eta_2\pm\sqrt{(1+\eta_1-\eta_2)^2-4\eta_1}\right],
\end{eqnarray}
with the factor $\eta_{1,2}=\frac{\omega_{1,2}^2}{m^2_{B}}$.
If the meson pairs are in the $P$-wave configuration, the corresponding longitudinal polarization vectors are defined as
 \begin{eqnarray}\label{eq:pq1}
\epsilon_{p}=\frac{1}{\sqrt{2\eta_1}}(g^+,-g^-,\textbf{0}_{T}),\quad
\epsilon_{q}=\frac{1}{\sqrt{2\eta_2}}(-f^-,f^+,\textbf{0}_{T}),
\end{eqnarray}
which satisfy the normalization $\epsilon_{p}^2=\epsilon_{q}^2=-1$  and the orthogonality
$\epsilon_{p}\cdot p=\epsilon_{q}\cdot q=0$.

As usual we also define the momentum $p_{1,2}$ of pion-pion pair and $q_{1,2}$ for the kaon-pion pair as
\begin{eqnarray}\label{eq:p1p2}
p_1&=&\left(\frac{m_{B}}{\sqrt{2}}(\zeta_1+\frac{r_1-r_2}{2\eta_1})g^+,\frac{m_{B}}{\sqrt{2}}(1-\zeta_1+\frac{r_1-r_2}{2\eta_1})g^-,\textbf{p}_{T}\right),\non
p_2&=&\left(\frac{m_{B}}{\sqrt{2}}(1-\zeta_1-\frac{r_1-r_2}{2\eta_1})g^+,\frac{m_{B}}{\sqrt{2}}(\zeta_1-\frac{r_1-r_2}{2\eta_1})g^-,-\textbf{p}_{T}\right),\non
q_1&=&\left(\frac{m_{B}}{\sqrt{2}}(1-\zeta_2+\frac{r_3-r_4}{2\eta_2})f^-,\frac{m_{B}}{\sqrt{2}}(\zeta_2+\frac{r_3-r_4}{2\eta_2})f^+,\textbf{q}_{T}\right),\non
q_2&=&\left(\frac{m_{B}}{\sqrt{2}}(\zeta_2-\frac{r_3-r_4}{2\eta_2})f^-,\frac{m_{B}}{\sqrt{2}}(1-\zeta_2-\frac{r_3-r_4}{2\eta_2})f^+,-\textbf{q}_{T}\right),\non
p_{\rm T}^2&=&\zeta_1(1-\zeta_1)\omega_1^2+\frac{(m_1^2-m_2^2)^2}{4\omega_1^2}-\frac{m^2_1+m^2_2}{2},\non
q_{\rm T}^2&=&\zeta_2(1-\zeta_2)\omega_2^2+\frac{(m_3^2-m_4^2)^2}{4\omega_2^2}-\frac{m^2_3+m^2_4}{2},
\end{eqnarray}
with $\zeta_1+\frac{r_1-r_2}{2\eta_1}=p_1^+/p^+$ and $\zeta_2+\frac{r_3-r_4}{2\eta_2}=q_1^-/q^-$ characterizing the momentum fraction for one of pion-pion (kaon-pion) pair, and the mass ratios $r_{1,2,3,4}=m_{1,2,3,4}^2/m^2_B$, $m_{1,2,3,4}$ being the masses of the final state meson.

One can obtain the relation between $\zeta_{1,2}$ and the polar angle $\theta_{1,2}$ in the dimeson rest frame in Fig.~\ref{fig1},
\begin{eqnarray}\label{eq:cos}
2\zeta_{1}-1=\sqrt{1-2\frac{r_1+r_2}{\eta_1}+\frac{(r_1-r_2)^2}{\eta_1^2}}\cos\theta_1,\quad
2\zeta_{2}-1=\sqrt{1-2\frac{r_3+r_4}{\eta_2}+\frac{(r_3-r_4)^2}{\eta_2^2}}\cos\theta_2,
\end{eqnarray}
with the upper and lower limits of $\zeta_{1,2}$
\begin{eqnarray}
\zeta_{1\text{max,min}}=\frac{1}{2}\left[1\pm\sqrt{1+4\alpha_1}\right],~\quad
\zeta_{2\text{max,min}}=\frac{1}{2}\left[1\pm\sqrt{1+4\alpha_2}\right].
\end{eqnarray}
For the sake of simplicity, we generally use the factor $\alpha_1=-\frac{r_1+r_2}{2\eta_1}+\frac{(r_1-r_2)^2}{4\eta_1^2}$ and $\alpha_2=-\frac{r_3+r_4}{2\eta_2}+\frac{(r_3-r_4)^2}{4\eta_2^2}$ in the following sections.

The differential branching fraction for the $B_{(s)}\rightarrow (\pi\pi)(K\pi)$
in the $B_{(s)}$ meson rest frame is expressed as
\begin{eqnarray}\label{eq:decayrate}
\frac{d^5\mathcal{B}}{d\Omega}=\frac{\tau_B k(\omega_1)k(\omega_2)k(\omega_1,\omega_2)}{16(2\pi)^6m_B^2} |A|^2, 
\end{eqnarray}
where $d \Omega$ with $\Omega\equiv \{\theta_1, \theta_2,\phi,\omega_1,\omega_2\}$ stands for the
five-dimensional measure spanned by the three helicity angles and the two invariant masses, and
\begin{eqnarray}
k(\omega_1,\omega_2)&=&\frac{\sqrt{[m_B^2-(\omega_1+\omega_2)^2][m_B^2-(\omega_1-\omega_2)^2]}}{2m_B},
\end{eqnarray}
is the momentum of the $\pi\pi$ pair in the $B_{(s)}$ meson rest frame. The explicit expression of kinematic variables $k(\omega)$ is defined in the $h_1h_2$ center-of-mass frame
\begin{eqnarray}
k(\omega)=\frac{\sqrt{\lambda(\omega^2,m_{h_1}^2,m_{h_2}^2)}}{2\omega},
\end{eqnarray}
with the K$\ddot{a}$ll$\acute{e}$n function $\lambda (a,b,c)= a^2+b^2+c^2-2(ab+ac+bc)$ and $m_{h_1,h_2}$ being the final state mass.

The four-body phase space has been derived in the analyses of the
$K\rightarrow \pi\pi l\nu$ decay~\cite{pr168-1858}, the
semileptonic $\bar{B}\rightarrow D(D^*)\pi l \nu$ decays~\cite{prd48-3204},
semileptonic baryonic decays~\cite{prd85-094019,plb780-100}, and four-body baryonic decays~\cite{plb770-348}.
One can confirm that Eq.~(\ref{eq:decayrate}) is equivalent to those in Refs.~\cite{prd85-094019,plb770-348}
by appropriate variable changes.
Replacing the helicity angle $\theta$ by the meson momentum fraction $\zeta$ via Eq.~(\ref{eq:cos}),
the Eq.~(\ref{eq:decayrate}) is turned into
\begin{eqnarray}\label{eq:decayrate1}
\frac{d^5\mathcal{B}}{d\zeta_1d\zeta_2d \omega_1d \omega_2d\phi}=
\frac{\tau_B k(\omega_1)k(\omega_2)k(\omega_1,\omega_2)}{4(2\pi)^6m_B^2\sqrt{1+4\alpha_1}\sqrt{1+4\alpha_2}}|A|^2.
\end{eqnarray}

\subsection{Helicity amplitudes}
One can disentangle the helicities of $R_1(\to \pi\pi)R_2(\to K\pi)$ final state via an angular analysis, depicted in Fig.~\ref{fig1}.
Taking the $B^0 \to R_1R_2 \to (\pi^+\pi^-)(K^+\pi^-)$ decay as an example, the $\theta_1$ is the angle between the $\pi^+$ direction in the $(\pi^+\pi^-)$ rest frame and the $(\pi^+\pi^-)$ direction in the $B^0$ rest frame, $\theta_2$ is the angle between the $K^+$ direction in the $(K^+\pi^-)$ rest frame and the $(K^+\pi^-)$ direction in the $B^0$ rest frame, and $\phi$ is the angle between the $(\pi^+\pi^-)$ and $(K^+\pi^-)$ decay planes.
A $\pi\pi(K\pi)$ pair can be produced in the $S$ or $P$-wave configuration in the selected invariant mass regions.

One decomposes the decay amplitudes into six helicity components: $h=VV$ (3), $VS$, $SV$, and $SS$, each with a corresponding amplitude $A_h$.
The first three, commonly referred to as the $P$-wave amplitudes, are associated with
the final states, where both $\pi\pi$ and $K\pi$ pairs come from intermediate vector mesons.
In the transversity basis, a $P$-wave decay amplitude can be decomposed into three components:
$A_0$, for which the polarizations of the final-state vector mesons are
longitudinal to their momenta, and $A_{\parallel}$ ($A_\perp$), for which the polarizations
are transverse to the momenta and parallel (perpendicular) to each other.
As the $S$-wave $\pi\pi(K\pi)$ pair arises from $R_1$ ($R_2$) labelled in Fig.~\ref{fig2}(a),
the corresponding single $S$-wave amplitude is denoted $A_{SV}$ ($A_{VS}$).
The double $S$-wave amplitude $A_{SS}$ is associated with the final state,
where both two-meson pairs are generated in the $S$ wave.
Specifically, these helicity amplitudes for the $B_{(s)}\to (\pi\pi)(K\pi)$ decay
denote
\begin{eqnarray}
A_{VV}&:& B_{(s)} \rightarrow \rho(\to \pi\pi) K^*(\to K\pi), \nonumber\\
A_{VS}&:& B_{(s)} \rightarrow \rho(\to \pi\pi)K_0^*(1430)(\to K\pi) ,\nonumber\\
A_{SV}&:& B_{(s)} \rightarrow f_0(500,980)(\to \pi\pi) K^*(\to K\pi),\nonumber\\
A_{SS}&:& B_{(s)} \rightarrow f_0(500,980)(\to \pi\pi) K_0^*(1430)(\to K\pi).
\end{eqnarray}

Relying on the Eq.~(\ref{eq:cos}), we get the full decay amplitude in Eq.~(\ref{eq:decayrate1}) as a coherent sum of the $P$-, $S$-, and double $S$-wave components by including the $\zeta_{1,2}$ dependencies instead of $\theta_{1,2}$ and azimuth-angle dependencies,
\begin{eqnarray}\label{eq:allampli}
A&=&A_0\frac{2\zeta_1-1}{\sqrt{1+4\alpha_1}}\frac{2\zeta_2-1}{\sqrt{1+4\alpha_2}}
+A_{\parallel}2\sqrt{2}\sqrt{\frac{\zeta_1(1-\zeta_1)+\alpha_1}{1+4\alpha_1}}
\sqrt{\frac{\zeta_2(1-\zeta_2)+\alpha_2}{1+4\alpha_2}}\cos\phi
\nonumber\\&&
+i A_{\perp}2\sqrt{2}\sqrt{\frac{\zeta_1(1-\zeta_1)+\alpha_1}{1+4\alpha_1}}
\sqrt{\frac{\zeta_2(1-\zeta_2)+\alpha_2}{1+4\alpha_2}}\sin\phi
+A_{VS}\frac{2\zeta_1-1}{\sqrt{1+4\alpha_1}}+A_{SV}\frac{2\zeta_2-1}{\sqrt{1+4\alpha_2}}+A_{SS}.
\end{eqnarray}

On basis of Eq.~(\ref{eq:decayrate1}), we can obtain the branching ratio for each helicity state,
\begin{eqnarray}\label{eq:brsss}
\mathcal{B}_h=\frac{\tau_B}{4(2\pi)^6m_B^2}\frac{2\pi}{9}C_h
\int d\omega_1d\omega_2k(\omega_1)k(\omega_2)k(\omega_1,\omega_2)|A_h|^2,
\end{eqnarray}
where the invariant masses $\omega_{1,2}$ are integrated over the chosen $\pi\pi$ and $K\pi$ mass window, respectively.
The coefficients $C_h$ are the results of the integrations over $\zeta_1,\zeta_2,\phi$ in terms of Eq.~(\ref{eq:brsss}) and listed as follows,
\begin{eqnarray}\label{eq:radii}
C_h=\left\{
\begin{aligned}
&(1+4\alpha_1)(1+4\alpha_2) \quad\quad\quad  &h=0,\parallel,\perp \\
&3(1+4\alpha_{1,2}) \quad\quad\quad  &h=VS,SV \\
&9 \quad\quad\quad  &h=SS. \\
\end{aligned}\right.
\end{eqnarray}

The $CP$-averaged branching ratio and the direct $CP$ asymmetry in each component are defined as below,
\begin{eqnarray}
\mathcal{B}_h^{avg}=\frac{1}{2}(\mathcal{B}_h+\mathcal{\bar{B}}_h),
\quad \mathcal{A}^{\text{dir}}_h=\frac{\mathcal{\bar{B}}_h-\mathcal{B}_h}{\mathcal{\bar{B}}_h+\mathcal{B}_h},
\end{eqnarray}
respectively, where $\mathcal{\bar{B}}_h$ is the branching ratio of the corresponding $CP$-conjugate channel.

For the $VV$ decays, the polarization fractions $f_{\lambda}$ with $\lambda=0$, $\parallel$,
and $\perp$ are described as
\begin{eqnarray}\label{pol}
f_{\lambda}=\frac{\mathcal{B}_{\lambda}}{\mathcal{B}_0+\mathcal{B}_{\parallel}+\mathcal{B}_{\perp}},
\end{eqnarray}
with the normalisation relation $f_0+f_{\parallel}+f_{\perp}=1$.
\subsection{Triple product asymmetries}\label{sec:TPAs}
In this work, we not only calculate the direct $CP$ asymmetries, but also pay more attention to the TPAs.
Consider a four-body decay $B \to R_1 (\to P_1P_2) R_2(\to Q_1Q_2)$, in which one measures the four particles' momenta in the $B$ rest frame.
We define $\hat{n}_{R_i} (i=1,2)$ is a unit vector perpendicular to the $R_i$ decay plane and $\hat{z}$ is a unit vector in the direction of $R_1$ in the $B$ rest frame.
Thus we have
\begin{eqnarray}\label{TP1}
\hat{n}_{R_1}\cdot \hat{n}_{R_2}=\cos\phi, ~\quad \hat{n}_{R_1}\times \hat{n}_{R_2}=\sin\phi \hat{z},
\end{eqnarray}
implying a $T$-odd scalar triple product
\begin{eqnarray}\label{TP2}
(\hat{n}_{R_1}\times \hat{n}_{R_2})\cdot \hat{z}=\sin\phi, ~\quad 2(\hat{n}_{R_1}\cdot \hat{n}_{R_2})(\hat{n}_{R_1}\times \hat{n}_{R_2})\cdot\hat{z}=\sin2\phi .
\end{eqnarray}

One can define a TPA as an asymmetry between the number of decays involving positive and negative values of $\sin\phi$ or $\sin2\phi$,
\begin{eqnarray} \label{eq:ATs}
\mathcal{A}_T^1&=&\frac{\Gamma(\cos\theta_1\cos\theta_2\sin\phi>0)-\Gamma(\cos\theta_1\cos\theta_2\sin\phi<0)}
{\Gamma(\cos\theta_1\cos\theta_2\sin\phi>0)+\Gamma(\cos\theta_1\cos\theta_2\sin\phi<0)},\\
\mathcal{A}_T^2&=&\frac{\Gamma(\sin2\phi>0)-\Gamma(\sin2\phi<0)}
{\Gamma(\sin2\phi>0)+\Gamma(\sin2\phi<0)}.
\end{eqnarray}

According to Eq.~(\ref{eq:cos}), the TPAs associated with $A_{\perp}$ for the considered four-body decays are derived from the
partially integrated differential decay rates as~\cite{prd84-096013,jhep07-166}
\begin{eqnarray} \label{eq:ATs2}
\mathcal{A}_T^1&=&\frac{\Gamma((2\zeta_1-1)(2\zeta_2-1)\sin\phi>0)-\Gamma((2\zeta_1-1)(2\zeta_2-1)\sin\phi<0)}
{\Gamma((2\zeta_1-1)(2\zeta_2-1)\sin\phi>0)+\Gamma((2\zeta_1-1)(2\zeta_2-1)\sin\phi<0)}\nonumber\\&=&
-\frac{2\sqrt{2}}{\pi\mathcal{D}}\int d\omega_1 d\omega_2k(\omega_1)k(\omega_2)k(\omega_1,\omega_2) \text{Im}[A_{\perp}A_0^*],\\
\mathcal{A}_T^2&=&\frac{\Gamma(\sin2\phi>0)-\Gamma(\sin2\phi<0)}
{\Gamma(\sin2\phi>0)+\Gamma(\sin2\phi<0)}\nonumber\\&=&
-\frac{4}{\pi\mathcal{D}}\int d\omega_1 d\omega_2k(\omega_1)k(\omega_2)k(\omega_1,\omega_2) \text{Im}[A_{\perp}A_{\parallel}^*],
\end{eqnarray}
with the denominator
\begin{eqnarray}
\mathcal{D}=\int d\omega_1 d\omega_2 k(\omega_1)k(\omega_2)k(\omega_1,\omega_2)(|A_0|^2+|A_{\parallel}|^2+|A_{\perp}|^2).
\end{eqnarray}

It has been found that TPAs originate from the interference of the $CP$-odd amplitudes $A_{\perp}$ with the other $CP$-even amplitudes $A_0$, $A_{\parallel}$.
The above TPAs contain the integrands $\text{Im}[A_{\perp}A_i^*]=|A_{\perp}||A_i^*|\sin(\Delta\phi+\Delta \delta)$ with $i={0,\parallel}$,
where $\Delta\phi$ and $\Delta\delta$ denote the weak and strong phase differences between the amplitudes
$A_{\perp}$ and $A_i$, respectively.
As already noted, $\text{Im}[A_{\perp}A_i^*]$ can be nonzero even if the weak phases vanish.
Thus, it is not quite accurate to identify a nonzero TPA as a signal of $CP$ violation.
To obtain a true $CP$ violation signal, one has to compare the TPAs in the $B$ and $\bar{B}$ meson decays.
The helicity amplitude for the $CP$-conjugated process can be inferred from  Eq.~(\ref{eq:allampli})
through $A_0\to \bar{A}_0$, $A_{\parallel} \to \bar{A}_{\parallel}$ and $A_{\perp} \to -\bar{A}_{\perp}$, in which the $\bar{A}_{\lambda}$ are obtained from the $A_{\lambda}$ by changing the sign of the weak phases.
Thus, the TPAs $\bar{\mathcal{A}}_T^i$ for the charge-conjugate process are defined similarly,
but with a multiplicative minus sign.
We denote by $\bar{A}_0,\bar{A}_{\parallel},\bar{A}_{\perp}$ transversity amplitudes for the $CP$-conjugate decay $\bar{B}\to \bar{R}_1(\to \bar{P}_1\bar{P}_2)\bar{R}_2(\to \bar{Q}_1\bar{Q}_2)$ and the corresponding three angles will be denoted by $\bar{\theta}_1$, $\bar{\theta}_2$, and $\bar{\phi}$.
Obviously $\theta_1=\bar{\theta}_1$, $\theta_2=\bar{\theta}_2$ and $\phi=\bar{\phi}$.
We then have the TPAs for the $CP$-averaged decay rates~\cite{prd84-096013}
\begin{eqnarray}\label{eq:tpa}
\mathcal{A}_T^{(1)\text{ave}}(\text{true})&\equiv& \frac{[\Gamma(T>0)+\bar{\Gamma}(\bar{T}>0)]-[\Gamma(T<0)+\bar{\Gamma}(\bar{T}<0)]}
{[\Gamma(T>0)+\bar{\Gamma}(\bar{T}>0)]+[\Gamma(T<0)+\bar{\Gamma}(\bar{T}<0)]}\non
&=&-\frac{2\sqrt{2}}{\pi(\mathcal{D}+\bar{\mathcal{D}})}\int d\omega_1 d\omega_2k(\omega_1)k(\omega_2)k(\omega_1,\omega_2) \text{Im}[A_{\perp}A_0^*-\bar{A}_{\perp}\bar{A}_0^*], \label{tp1-t}\\
\mathcal{A}_T^{(2)\text{ave}}(\text{true})&\equiv& \frac{[\Gamma(\sin2\phi>0)+\bar{\Gamma}(\sin2\bar{\phi}>0)]-[\Gamma(\sin2\phi<0)+\bar{\Gamma}(\sin2\bar{\phi}<0)]}
{[\Gamma(\sin2\phi>0)+\bar{\Gamma}(\sin2\bar{\phi}>0)]+[\Gamma(\sin2\phi<0)+\bar{\Gamma}(\sin2\bar{\phi}<0)]}\non
&=&-\frac{4}{\pi(\mathcal{D}+\bar{\mathcal{D}})}\int d\omega_1 d\omega_2k(\omega_1)k(\omega_2)k(\omega_1,\omega_2) \text{Im}[A_{\perp}A_{\parallel}^*-\bar{A}_{\perp}\bar{A}_{\parallel}^*], \label{tp2-t}\\
\mathcal{A}_T^{(1)\text{ave}}(\text{fake})&\equiv& \frac{[\Gamma(T>0)-\bar{\Gamma}(\bar{T}>0)]-[\Gamma(T<0)-\bar{\Gamma}(\bar{T}<0)]}
{[\Gamma(T>0)+\bar{\Gamma}(\bar{T}>0)]+[\Gamma(T<0)+\bar{\Gamma}(\bar{T}<0)]}\non
&=&-\frac{2\sqrt{2}}{\pi(\mathcal{D}+\bar{\mathcal{D}})}\int d\omega_1 d\omega_2k(\omega_1)k(\omega_2)k(\omega_1,\omega_2) \text{Im}[A_{\perp}A_0^*+\bar{A}_{\perp}\bar{A}_0^*], \label{tp1-f}\\
\mathcal{A}_T^{(2)\text{ave}}(\text{fake})&\equiv& \frac{[\Gamma(\sin2\phi>0)-\bar{\Gamma}(\sin2\bar{\phi}>0)]-[\Gamma(\sin2\phi<0)-\bar{\Gamma}(\sin2\bar{\phi}<0)]}
{[\Gamma(\sin2\phi>0)+\bar{\Gamma}(\sin2\bar{\phi}>0)]+[\Gamma(\sin2\phi<0)+\bar{\Gamma}(\sin2\bar{\phi}<0)]}\non
&=&-\frac{4}{\pi(\mathcal{D}+\bar{\mathcal{D}})}\int d\omega_1 d\omega_2k(\omega_1)k(\omega_2)k(\omega_1,\omega_2) \text{Im}[A_{\perp}A_{\parallel}^*+\bar{A}_{\perp}\bar{A}_{\parallel}^*], \label{tp2-f}
\end{eqnarray}
with $\bar{\Gamma}$ being the decay rate of the $CP$-conjugate process, $T=(2\zeta_1-1)(2\zeta_2-1)\sin\phi$ and $\bar{T}=(2\zeta_1-1)(2\zeta_2-1)\sin\bar{\phi}$ and the denominator is
\begin{eqnarray}
\bar{\mathcal{D}}=\int d\omega_1 d\omega_2 k(\omega_1)k(\omega_2)k(\omega_1,\omega_2)(|\bar{A}_0|^2+|\bar{A}_{\parallel}|^2+|\bar{A}_{\perp}|^2),
\end{eqnarray}
for the $CP$-conjugate decay.

It is shown that the term $\text{Im}[A_{\perp}A_{0,\parallel}^*-\bar{A}_{\perp}\bar{A}_{0,\parallel}^*]$ is proportional to $\sin\Delta\phi\cos\Delta\delta$, which is nonzero only in the presence of the weak phase difference.
Then TPAs provide an alternative measure of $CP$ violation.
Furthermore, compared with direct $CP$ asymmetries, $\mathcal{A}_T^{(i)\text{ave}}(\text{true})$ does not suffer the suppression from the strong phase difference, and is maximal when the strong phase difference vanishes.
While for the term $\text{Im}[A_{\perp}A_{0,\parallel}^*+\bar{A}_{\perp}\bar{A}_{0,\parallel}^*]\propto\cos\Delta\phi\sin\Delta\delta$,
the $\mathcal{A}_T^{(i)\text{ave}}(\text{fake})$ can be nonzero when the weak phase difference vanishes.
Such a quantity is referred as a fake asymmetry ($CP$ conserving), which
reflects the effect of strong phases~\cite{prd84-096013,plb701-357}, instead of $CP$ violation.

For a more direct comparison with the measurements from LHCb~\cite{jhep05-026}, the so-called $true$ and $fake$ TPAs are then defined as
\begin{eqnarray}\label{eq:ture1}
\mathcal{A}_T^1(\text{true})&\equiv&\frac{1}{2}(\mathcal{A}_T^1+\bar{\mathcal{A}}_T^1)=-\frac{\sqrt{2}}{\pi}\int d\omega_1 d\omega_2k(\omega_1)k(\omega_2)k(\omega_1,\omega_2) \text{Im}[\frac{A_{\perp}A_0^*}{\mathcal{D}}-\frac{\bar{A}_{\perp}\bar{A}_0^*}{\bar{\mathcal{D}}}],\label{tp1p-t}\\
\mathcal{A}_T^2(\text{true})&\equiv&\frac{1}{2}(\mathcal{A}_T^2+\bar{\mathcal{A}}_T^2)
=-\frac{2}{\pi}\int d\omega_1 d\omega_2k(\omega_1)k(\omega_2)k(\omega_1,\omega_2) \text{Im}[\frac{A_{\perp}A_{\parallel}^*}{\mathcal{D}}-\frac{\bar{A}_{\perp}\bar{A}_{\parallel}^*}{\bar{\mathcal{D}}}], \label{tp2p-t}\\
\mathcal{A}_T^1(\text{fake})&\equiv&\frac{1}{2}(\mathcal{A}_T^1-\bar{\mathcal{A}}_T^1)=-\frac{\sqrt{2}}{\pi}\int d\omega_1 d\omega_2k(\omega_1)k(\omega_2)k(\omega_1,\omega_2) \text{Im}[\frac{A_{\perp}A_0^*}{\mathcal{D}}+\frac{\bar{A}_{\perp}\bar{A}_0^*}{\bar{\mathcal{D}}}], \label{tp1p-f}\\
\mathcal{A}_T^2(\text{fake})&\equiv&\frac{1}{2}(\mathcal{A}_T^2-\bar{\mathcal{A}}_T^2)
=-\frac{2}{\pi}\int d\omega_1 d\omega_2k(\omega_1)k(\omega_2)k(\omega_1,\omega_2) \text{Im}[\frac{A_{\perp}A_{\parallel}^*}{\mathcal{D}}+\frac{\bar{A}_{\perp}\bar{A}_{\parallel}^*}{\bar{\mathcal{D}}}], \label{tp2p-f}
\end{eqnarray}
where the $true$ or $fake$ labels refer to whether the asymmetry is due to a real $CP$ asymmetry or due to effects from final-state interactions that are $CP$ symmetric.
It should be stressed that two asymmetries defined in Eqs.~(\ref{tp1-t}) and (\ref{tp1p-t}) are different in the most case, as well as two asymmetries in Eqs.~(\ref{tp2-t}) and (\ref{tp2p-t}).
They become equal when no direct $CP$ violation occurs in the total rate, namely $\mathcal{D}=\bar{\mathcal{D}}$.

\subsection{Two-meson distribution amplitudes}
\subsubsection{$S$-wave two-meson DAs}
Here we briefly introduce the $S$ and $P$-wave two-meson DAs and the corresponding time-like form factors used in our framework.
One can see that resonant contributions through two-body channels can be included by parameterizing the two-meson DAs.
The $S$-wave two-meson DA is written in the following form~\cite{prd91-094024},
\begin{eqnarray}\label{swave}
\Phi_{S}(z,\omega)=\frac{1}{\sqrt{2N_c}}[{p\hspace{-1.5truemm}/}\phi^0_S(z,\omega^2)+
\omega\phi^s_S(z,\omega^2)+\omega({n\hspace{-2.0truemm}/}{v\hspace{-2.0truemm}/}-1)\phi^t_S(z,\omega^2)].
\end{eqnarray}
In what follows the subscripts $S$ and $P$ are always associated with the corresponding partial waves.

For the scalar resonances $f_0(500)$ and $f_0(980)$, the asymptotic forms of the individual DAs in Eq.~(\ref{swave}) have been parameterized as~\cite{MP,MT01,MT02,MT03}
\begin{eqnarray}
\phi^0_S(z,\omega^2)&=&\frac{9F_S(\omega^2)}{\sqrt{2N_c}}a_S z(1-z)(1-2z),\\
\phi^s_S(z,\omega^2)&=&\frac{F_S(\omega^2)}{2\sqrt{2N_c}},\\
\phi^t_S(z,\omega^2)&=&\frac{F_S(\omega^2)}{2\sqrt{2N_c}}(1-2z),
\end{eqnarray}
with the time-like scalar form factor $F_S(\omega^2)$ and the Gegenbauer coefficient $a_S$.
While for the scalar resonance $K_0^*(1430)$, we will adopt similar formulas as those for a scalar meson~\cite{plb730-336,prd73-014017}, bearing in mind large uncertainties that may be introduced by this approximation.
The detailed expressions of DAs for various twists are as follows:
\begin{eqnarray}\label{eq:phi0st}
\phi^0_S&=&\frac{6}{2\sqrt{2N_c}}F_S(\omega^2)z(1-z)
\left [ \frac{1}{\mu_S}+B_1C_1^{3/2}(t)+B_3C_3^{3/2}(t) \right ], \label{eq:phi1s}\\
\phi^s_S&=&\frac{1}{2\sqrt{2N_c}}F_S(\omega^2),\label{eq:phi2s}\\
\phi^t_S&=&\frac{1}{2\sqrt{2N_c}}F_S(\omega^2)(1-2z)\label{eq:phi3s},
\end{eqnarray}
where the Gegenbauer polynomials $C_{1}^{3/2}(t)=3t$, $C_{3}^{3/2}(t)=\frac{5}{2}t(7t^2-3)$ with $t=1-2z$ and $\mu_S=\omega/(m_{02}-m_{01})$.
$\omega$ is the $K\pi$ invariant mass and $m_{01,02}$ are the running current quark masses.
The Gegenbauer moments $B_{1,3}$ at the 1 GeV scale from Scenario I in the QCD sum rule analysis and the related running current quark masses can be found in Refs.~\cite{prd73-014017,prd77-014034,prd78-014006}.

\subsubsection{$P$-wave two-meson DAs}
The $P$-wave two-meson DAs related to both longitudinal and transverse polarizations are organized in analogy with those in Ref.~\cite{prd98-113003}.
The explicit expressions of the $P$-wave pion-pion (kaon-pion) DAs associated with longitudinal and transverse polarization are described as follows,
\begin{eqnarray}
\Phi_P^{L}(z,\zeta,\omega)&=&\frac{1}{\sqrt{2N_c}} \left [{ \omega \epsilon\hspace{-1.5truemm}/_p  }\phi_P^0(z,\omega^2)+\omega\phi_P^s(z,\omega^2)
+\frac{{p\hspace{-1.5truemm}/}_1{p\hspace{-1.5truemm}/}_2
  -{p\hspace{-1.5truemm}/}_2{p\hspace{-1.5truemm}/}_1}{\omega(2\zeta-1)}\phi_P^t(z,\omega^2) \right ] (2\zeta-1)\;,\label{pwavel}\\
\Phi_P^{T}(z,\zeta,\omega)&=&\frac{1}{\sqrt{2N_c}}
\Big [\gamma_5{\epsilon\hspace{-1.5truemm}/}_{T}{ p \hspace{-1.5truemm}/ } \phi_P^T(z,\omega^2)
+\omega \gamma_5{\epsilon\hspace{-1.5truemm}/}_{T} \phi_P^a(z,\omega^2)+ i\omega\frac{\epsilon^{\mu\nu\rho\sigma}\gamma_{\mu}
\epsilon_{T\nu}p_{\rho}n_{-\sigma}}{p\cdot n_-} \phi_P^v(z,\omega^2) \Big ]\sqrt{\zeta(1-\zeta)+\alpha_1}\label{pwavet}\;.
\end{eqnarray}

The two-pion DAs for various twists are expanded in terms of the Gegenbauer polynomials:
\begin{eqnarray}
\phi_P^0(z,\omega^2)&=&\frac{3F_P^{\parallel}(\omega^2)}{\sqrt{2N_c}}z(1-z)\left[1
+a^0_{2\rho}\frac{3}{2}(5(1-2z)^2-1)\right] \;,\\
\phi_P^s(z,\omega^2)&=&\frac{3F_P^{\perp}(\omega^2)}{2\sqrt{2N_c}}(1-2z)\left[1
+a^s_{2\rho}(10z^2-10z+1)\right]  \;,\\
\phi_P^t(z,\omega^2)&=&\frac{3F_P^{\perp}(\omega^2)}{2\sqrt{2N_c}}(1-2z)^2\left[1
+a^t_{2\rho}\frac{3}{2}(5(1-2z)^2-1)\right]  \;,\\
\phi_P^T(z,\omega^2)&=&\frac{3F_P^{\perp}(\omega^2)}
{\sqrt{2N_c}}z(1-z)[1+a^{T}_{2\rho}\frac{3}{2}(5(1-2z)^2-1)]\;,\\
\phi_P^a(z,\omega^2)&=&\frac{3F_P^{\parallel}(\omega^2)}
{4\sqrt{2N_c}}(1-2z)[1+a_{2\rho}^a(10z^2-10z+1)]\;,\\
\phi_P^v(z,\omega^2)&=&\frac{3F_P^{\parallel}(\omega^2)}
{8\sqrt{2N_c}}\bigg\{[1+(1-2z)^2]+a^v_{2\rho}[3(2z-1)^2-1]\bigg\}\;,
\end{eqnarray}
with the Gegenbauer coefficients $a_{2\rho}^i$ and the two $P$-wave form factors $F_P^{\parallel}(\omega^2)$ and $F_P^{\perp}(\omega^2)$.

For the kaon-pion DAs, the various twist DAs $\phi_P^i$ have similar forms as the corresponding ones for the $K^*$ meson~\cite{prd71-114008} by replacing the
decay constants with the time-like form factors,
\begin{eqnarray}
\phi_P^0(z,\omega^2)&=&\frac{3F_P^{\parallel}(\omega^2)}{\sqrt{2N_c}} z(1-z)\left[1+a_{1K^*}^{||}3t+a_{2K^*}^{||}\frac{3}{2}(5t^2-1)\right]\;,\label{eqphi0}\\
\phi_P^s(z,\omega^2)&=&\frac{3F_P^{\perp}(\omega^2)}{2\sqrt{2N_c}}t\;,\label{eqphis}\\
\phi_P^t(z,\omega^2)&=&\frac{3F_P^{\perp}(\omega^2)}{2\sqrt{2N_c}} t^2\;,\label{eqphit}\\
\phi_P^T(z,\omega^2)&=&\frac{3F_P^{\perp}(\omega^2)}{\sqrt{2N_c}}z(1-z)\left[1+a_{1K^*}^{\perp}3 t+
a_{2K^*}^{\perp}\frac{3}{2}(5 t^2-1)\right]\;,\label{eqphi}\\
\phi_P^a(z,\omega^2)&=&\frac{3F_P^{\parallel}(\omega^2)}{4\sqrt{2N_c}}t\;,\label{eqphia}\\
\phi_P^v(z,\omega^2)&=&\frac{3F_P^{\parallel}(\omega^2)}{8\sqrt{2N_c}}(1+ t^2)\;,\label{eqphiv}
\end{eqnarray}
with $t=(1-2z)$.
The Gegenbauer moments associated with longitudinal polarization $a_{1K^*}^{\parallel},a_{2K^*}^{\parallel}$ are determined in Ref.~\cite{2105-03899}, while the Gegenbauer moments associated with transverse polarization $a_{1K^*}^{\perp},a_{2K^*}^{\perp}$ are adopted the same values as those longitudinal ones.
\subsubsection{Time-like form factor}
The strong interactions between the resonance and the final-state meson pair, including elastic rescattering of the final-state meson pair, can be factorized into the time-like form factor $F_{S,P}(\omega^2)$, which is guaranteed by the Watson theorem~\cite{pr88-1163}.
We usually use the relativistic Breit-Wigner (BW) line shape for a narrow resonance to parameterize the time-like form factors $F^{\parallel}(\omega^2)$~\cite{BW-model}.
The explicit formula is expressed as~\cite{epjc78-1019},
\begin{eqnarray}
\label{BRW}
F^{\parallel}(\omega^2)&=&\frac{ m_i^2}{m^2_i -\omega^2-im_i\Gamma_i(\omega^2)} \;,
\end{eqnarray}
where the $m_i$ and $\Gamma_i$ are the pole mass and width of the corresponding resonances shown in Table~\ref{Tab:pa}, respectively.
The mass-dependent width $\Gamma_i(\omega)$ is defined as
\begin{eqnarray}
\label{BRWl}
\Gamma_i(\omega^2)&=&\Gamma_i\left(\frac{m_i}{\omega}\right)\left(\frac{k(\omega)}{k(m_i)}\right)^{(2L_R+1)}\Blue{.}
\end{eqnarray}
The $k(\omega)$ is the momentum vector of the resonance decay product measured in the resonance rest frame, while $k(m_i)$ is the value of $k(\omega)$ when $\omega=m_i$.
$L_R$ is the orbital angular momentum in the $\pi\pi$ ($K\pi$) system and $L_R=0,1,...$ corresponds to the $S,P,...$ partial-wave resonances.
Due to the limited studies on the form factor $F^{\perp}(\omega^2)$, we use the two decay
constants $f_i^{(T)}$ of the intermediate particle to determine the ratio $F^{\perp}(\omega^2)/F^{\parallel}(\omega^2)\approx (f_i^T/f_i)$.

The BW formula does not work well for $f_0(980)$, since its pole mass is close to the $K\bar{K}$ threshold.
For scalar resonance $f_0(980)$, we adopt the Flatt\'e parametrization where the resulting line shape is above and below the threshold of the intermediate particle~\cite{plb63-228}.
If the coupling of a resonance to the channel opening nearby is very strong, the Flatt\'e parametrization shows a scaling invariance and does not allow for an extraction of individual partial decay widths.
Thus, we employ the modified Flatt\'e model suggested by Bugg~\cite{prd78-074023} following the LHCb collaboration~\cite{prd89-092006,prd90-012003},
\begin{eqnarray}
F(\omega^2)=\frac{m_{f_0(980)}^2}{m_{f_0(980)}^2-\omega^2-im_{f_0(980)}(g_{\pi\pi}\rho_{\pi\pi}+g_{KK}\rho_{KK}F^2_{KK})}\;.
\end{eqnarray}
The coupling constants $g_{\pi\pi}=0.167$ GeV and $g_{KK}=3.47g_{\pi\pi}$~\cite{prd89-092006,prd90-012003}
describe the $f_0(980)$ decay into the final states $\pi^+\pi^-$ and $K^+K^-$, respectively.
The phase space factors $\rho_{\pi\pi}$ and $\rho_{KK}$ read
as~\cite{prd87-052001,prd89-092006,plb63-228}
\begin{eqnarray}
\rho_{\pi\pi}=\frac23\sqrt{1-\frac{4m^2_{\pi^\pm}}{\omega^2}}
 +\frac13\sqrt{1-\frac{4m^2_{\pi^0}}{\omega^2}},\quad
\rho_{KK}=\frac12\sqrt{1-\frac{4m^2_{K^\pm}}{\omega^2}}
 +\frac12\sqrt{1-\frac{4m^2_{K^0}}{\omega^2}}.
\end{eqnarray}

If there are overlapping resonances or there is significant interference with a nonresonant component both in the same partial wave, the relativistic BW function leads to unitarity violation within the isobar model~\cite{0712-1605}.
This is the case for the $K\pi$ $S$-wave  at low $K\pi$ mass, where the $K^*_0(1430)$ resonance interferes strongly with a slowly varying NR $S$-wave component.
In this work, the time-like scalar form factor $F_S(\omega^2)$ for the $S$-wave $K\pi$ system is parametrized by using a modified LASS line shape~\cite{npb296-493} for the $S$-wave resonance $K^*_0(1430)$, which has been widely used in experimental analyses~\cite{epjc78-1019},
\begin{eqnarray}\label{eq:lass}
F_S(\omega^2)=\frac{\omega}{k(\omega)[\cot(\delta_B)-i]}+e^{2i\delta_B}\frac{m_0^2
\Gamma_0/k(m_0)}{m_0^2-\omega^2-im_0^2 \frac{\Gamma_0}{\omega }\frac{k(\omega)}{k(m_0)}}\;,
\end{eqnarray}
with
\begin{eqnarray}\label{eq:ar}
\cot(\delta_B)=\frac{1}{ak(\omega)}+\frac{rk(\omega)}{2},
\end{eqnarray}
where the first term in Eq.~(\ref{eq:lass}) is an empirical term from the elastic kaon-pion scattering and the second term is the resonant contribution with a phase factor to retain unitarity.
Here $m_0$ and $\Gamma_0$ are the pole mass and width of the $K^*_0(1430)$ state.
The parameters $a=(3.1\pm 1.0)~{\rm GeV^{-1}}$ and  $r=(7.0\pm 2.4)~{\rm GeV^{-1}}$ are the scattering length and effective range~\cite{epjc78-1019}, respectively.
The slowly varying part (the first term in the Eq.~(\ref{eq:lass})) is not well modeled at high masses and it is set to  zero for $m(K\pi)$ values above 1.7 GeV~\cite{epjc78-1019}.

In experimental investigations of three-body hadronic $B$ meson decays, the wide $\rho$ resonant contribution is usually parameterized as the Gounaris-Sakurai (GS)
model~\cite{prl21-244} based on the BW function~\cite{BW-model}.
It is a way to interpret the observed structures beyond the $\rho$ resonance in terms of higher mass isovector vector mesons.
By taking the $\rho-\omega$ interference and the excited states into account,
the form factor $F^{\parallel}(\omega^2)$ can be written in the form of~\cite{prd86-032013}
\begin{eqnarray}
F^{\parallel}(\omega^2)= \left [ {\rm GS}_\rho(s,m_{\rho},\Gamma_{\rho})
\frac{1+c_{\omega} {\rm BW}_{\omega}(s,m_{\omega},\Gamma_{\omega})}{1+c_{\omega}}
+\Sigma c_i {\rm GS}_i(s,m_i,\Gamma_i)\right] \left[ 1+\Sigma c_i\right]^{-1}\;,
\label{GS}
\end{eqnarray}
where $s=m^2(\pi\pi)$ is the two-pion invariant mass squared, $i=(\rho^{\prime}(1450), \rho^{\prime \prime}(1700), \rho^{\prime \prime \prime}(2254))$,
$\Gamma_{\rho,\omega,i}$ is the decay width for the relevant resonance, $m_{\rho,\omega,i}$ are the masses of the corresponding mesons, respectively.
The explicit expressions of the function ${\rm GS}_\rho(s,m_{\rho},\Gamma_{\rho})$ are described as follows~\cite{BW-model}
\begin{equation}
{\rm GS}_\rho(s, m_\rho, \Gamma_\rho) =
\frac{m_\rho^2 [ 1 + d(m_\rho) \Gamma_\rho/m_\rho ] }{m_\rho^2 - s + f(s, m_\rho, \Gamma_\rho)
- i m_\rho \Gamma (s, m_\rho, \Gamma_\rho)}~,
\end{equation}
with the functions
\begin{eqnarray}
\Gamma (s, m_\rho, \Gamma_\rho) &=& \Gamma_\rho  \frac{s}{m_\rho^2}
\left( \frac{\beta_\pi (s) }{ \beta_\pi (m_\rho^2) } \right) ^3~,\non
d(m) &=& \frac{3}{\pi} \frac{m_\pi^2}{k^2(m^2)} \ln \left( \frac{m+2 k(m^2)}{2 m_\pi} \right)
   + \frac{m}{2\pi  k(m^2)}
   - \frac{m_\pi^2  m}{\pi k^3(m^2)}~,\non
f(s, m, \Gamma) &=& \frac{\Gamma  m^2}{k^3(m^2)} \left[ k^2(s) [ h(s)-h(m^2) ]
+ (m^2-s) k^2(m^2)  h'(m^2)\right]~,\non
k(s) &=& \frac{1}{2} \sqrt{s}  \beta_\pi (s)~,\non
h(s) &=& \frac{2}{\pi}  \frac{k(s)}{\sqrt{s}}  \ln \left( \frac{\sqrt{s}+2 k(s)}{2 m_\pi} \right),
\end{eqnarray}
where $\beta_\pi (s) = \sqrt{1 - 4m_\pi^2/s}$.
Since the process $\omega \to \pi^+\pi^-$ suffers the G-parity suppression, we find that the interference between the $\rho$ and $\omega$ does not
significantly change the values in our calculations. Hence, it's reasonable to set $c_{\omega}=0$ in our latter numerical calculations. 

In contrast to the vector resonances, the identification of the scalar mesons is a long-standing puzzle.
Scalar resonances are difficult to resolve because some of them have large decay widths, which cause a strong overlap between resonances and background.
For a comparison, we here parameterize the $f_0(500)$ contribution in two different ways: the BW and the Bugg model~\cite{DVBugg:2007},
respectively.
The form factor of $f_0(500)$ with the Bugg resonant lineshape is written in the following form~\cite{DVBugg:2007}
\beq
T_{11}(s) &=& M \; \Gamma _1(s)   \left [ M^2 - s - g^2_1
\frac {s-s_A}{M^2-s_A} \left [ j_1(s) - j_1(M^2) \right ] - iM \sum_{i=1}^4 \Gamma_i(s) \right ]^{-1}, \label{eq:bugg}
\eeq
where $s=\omega^2 =m^2(\pi^+\pi^-)$, $j_1(s) = \frac {1}{\pi}\left[2 + \rho _1  \ln \left( \frac {1 - \rho _1}{1
+ \rho _1}\right) \right]$,
the functions $g^2_1(s)$, $\Gamma_i(s)$ and other relevant functions in Eq.~(\ref{eq:bugg} ) are the following
\beq
g^2_1(s) &=& M(b_1 + b_2s)\exp [-(s - M^2)/A],\non
M\; \Gamma_1(s) &=& g^2_1(s)\frac {s-s_A}{M^2-s_A}\rho_1(s), \non
M\; \Gamma_2(s) &=& 0.6g^2_1(s)(s/M^2)\exp (-\alpha |s-4m^2_K|)\rho_2(s),\non
M\;\Gamma_3(s) &=& 0.2g^2_1(s)(s/M^2)\exp (-\alpha |s-4m^2_\eta|)\rho_3(s), \non
M\;\Gamma_4(s) &=& M\; g_{4\pi}\; \rho_{4\pi}(s)/\rho_{4\pi }(M^2), \quad \rm{with} \quad
\rho _{4\pi}(s) = 1.0/[1 + \exp (7.082 - 2.845s)]. \label{eq:buggls}
\eeq
For the parameters in Eqs.~(\ref{eq:bugg},\ref{eq:buggls}), we use their values as given in the
fourth column of Table I in Ref.~\cite{DVBugg:2007}:
\beq
M &=& 0.953 {\rm GeV}, \quad s_A = 0.41 \ m_{\pi}^2, \quad b_1 = 1.302 {\rm GeV}^2,\non
b_2&=& 0.340, \quad A = 2.426 {\rm GeV}^2, \quad g_{4\pi} = 0.011 {\rm GeV}.
\label{eq:input1}
\eeq
And the parameters $\rho_{1,2,3}$ in Eq.~(\ref{eq:buggls})
are the phase-space factors of the decay channels $\pi\pi$, $KK$ and $\eta\eta$ respectively, and have been defined as
\cite{DVBugg:2007}
\beq
\rho_i(s) = \sqrt {1 - 4\frac{m^2_i }{s} },
\eeq
with $m_1=m_\pi, m_2=m_K$ and $m_3=m_\eta$.

\section{Numerical results}\label{sec:3}
In this section, we calculate the $CP$-averaged branching rations ($\cal B$), direct $CP$-violating asymmetries ($\cal A^{\text{CP}}$) and the polarization fractions $f_{\lambda}$, as well as estimate the size of TPAs, respectively.
The pole mass and decay width for the corresponding resonance have been summarized in Table~\ref{Tab:pa}.
Before proceeding with the numerical analysis, the meson masses and the decay constants  (in units of GeV), together with the $B$ meson lifetimes (in units of ps) are collected below~\cite{pdg2020,prd76-074018,prd95-056008}:
\begin{eqnarray}
m_{B}&=&5.280, \quad m_{B_s}=5.367, \quad m_b=4.8, \quad m_{K^\pm}=0.494, \non
m_{K^0}&=&0.498, \quad m_{\pi^{\pm}}=0.140, \quad m_{\pi^0}=0.135, \quad f_B=0.21\pm0.02, \non
f_{B_s}&=&0.23\pm0.02, \quad f_{\rho}=0.216 \pm 0.003, \quad f^T_{\rho}=0.184,\quad f_{K^*}=0.217 \pm 0.005, \non
f^T_{K^*}&=&0.185 \pm 0.010, \quad \tau_{B^0}=1.519,\quad \tau_{B^{\pm}}=1.638, \quad \tau_{B_{s}}=1.512.
\end{eqnarray}
The values of the Wolfenstein parameters are adopted as given in Ref.~\cite{pdg2018}:
$A=0.836\pm0.015, \lambda=0.22453\pm 0.00044$, $\bar{\rho} = 0.122^{+0.018}_{-0.017}$, $\bar{\eta}= 0.355^{+0.012}_{-0.011}$.

The Gegenbauer moments are adopted the same values as those determined
in Ref.~\cite{prd77-014034,prd98-113003,2105-03899,epjc76-675}
\begin{eqnarray}\label{eq:gen}
a_S&=&0.2\pm 0.2, \quad B_1=-0.57\pm0.13, \quad B_3=-0.42\pm0.22, \non
a^0_{2\rho}&=&0.08\pm0.13, \quad a^s_{2\rho}=-0.23\pm0.24,\quad a^t_{2\rho}=-0.354\pm0.062, \non
a^{T}_{2\rho}&=&0.50\pm0.50, \quad a^a_{2\rho}=0.40\pm0.40,\quad a^v_{2\rho}=-0.50\pm0.50,\non
a_{1K^*}^{||}&=&a_{1K^*}^{\perp}=0.31 \pm0.16, \quad a_{2K^*}^{||}=a_{2K^*}^{\perp}=1.188\pm0.098.
\end{eqnarray}

In our numerical calculations, the theoretical uncertainties quoted in the tables are estimated from three sources:
the first theoretical uncertainty results from the parameters of the wave functions of the initial states, such as the shape parameter $\omega_B=0.40\pm0.04$~GeV or $\omega_{B_s}=0.48\pm0.048$~GeV in $B_{(s)}$ meson wave function.
The second one is due to the Gegenbauer moments in various twist DAs of $\pi\pi$ and $K\pi$ pair with different intermediate resonances.
The last one is caused by the variation of the hard scale $t$ from $0.75t$ to $1.25t$ (without changing $1/b_i$) and the QCD scale $\Lambda_{\rm QCD}=0.25\pm0.05$~GeV, which characterizes the effect of the next-to-leading-order QCD contributions.
The possible errors due to the uncertainties of CKM matrix elements are very small and can be neglected safely.
The major uncertainty comes from the Gegenbauer moments, which amounts to $30\%\sim50\%$ in magnitude.
These parameters need to be constrained more precisely in order to improve the accuracy of theoretical predictions for four-body $B$ meson decays.

We perform an amplitude analysis of $B_{(s)} \to (\pi\pi)(K\pi)$ decays in the two-body invariant mass regions 300$<m(\pi\pi)<$1100 MeV, accounting for the $\rho$, $f_0(500)$ and $f_0(980)$ resonances, and 750$<m(K\pi)<$1200 MeV, which is dominated by the $K^*(892)$ meson, but the $S$-wave resonance $K_0^*(1430)$ is expected to contribute.
From the numerical results, one can address some issues as follows.

\begin{table}[t]
\caption{$CP$-averaged branching ratios and polarization fractions of the two-body $B_{(s)} \to \rho K^*$ decays in the PQCD approach compared with the previous predictions in the PQCD approach~\cite{prd91-054033}, the updated predictions in the QCDF~\cite{prd80-114026,prd80-114008},  SCET~\cite{prd96-073004} and FAT~\cite{epjc77-333}. Experimental results for branching ratios are taken from Table~\ref{exp1} and for polarization fractions from~\cite{1909-12524}.
The theoretical uncertainties are attributed to the variations of the shape parameter $\omega_{B_{(s)}}$ in the $B_{(s)}$ meson DA, of the Gegenbauer moments in various twist DAs of $\pi\pi$ and $K\pi$ pair, and of the hard scale $t$ and the QCD scale $\Lambda_{\rm QCD}$.}
\label{tab:br2body}
\begin{ruledtabular}
\begin{threeparttable}
\begin{tabular}[t]{lcccc}
Modes & ${\cal B} (10^{-6})$  &$f_0(\%)$ &$f_{\parallel}(\%)$&$f_{\perp}(\%)$ \\
\hline
$B^+ \to \rho^+ K^{*0}$ &$11.6^{+1.5+2.2+4.8}_{-1.2-2.3-3.5}$&$73.5^{+2.5+8.9+2.4}_{-2.3-9.4-3.3}$
                                &$13.4^{+1.2+4.7+1.1}_{-1.3-4.6-1.2}$&$13.1^{+1.1+4.7+2.2}_{-1.2-4.3-1.2}$\\
PQCD (former) &$9.9^{+4.7}_{-4.1}$&$70\pm5$&&$13.7^{+2.1}_{-1.9}$\\
QCDF &$9.2^{+3.8}_{-5.5}$&$48^{+52}_{-40}$&&\\
SCET &$8.93\pm3.18$&$45\pm18$&&$24.9\pm11.1$\\
FAT&$10.4\pm2.6$&$46.0\pm12.9$&&$27.2\pm7.0$\\
Expt.~\tnote{1}&$9.2\pm1.5$&$48\pm8$&&\\ \hline

$B^+ \to \rho^0 K^{*+}$ &$7.5^{+1.3+1.3+2.7}_{-0.9-1.2-2.2}$&$78.4^{+2.3+6.6+2.5}_{-2.1-7.1-3.4}$
                                &$13.3^{+1.0+3.9+2.3}_{-1.2-3.9-1.9}$&$8.3^{+1.1+3.3+1.2}_{-1.1-2.9-0.8}$\\
PQCD (former) &$6.1^{+2.8}_{-2.4}$&$75^{+4}_{-5}$&&$11.9^{+2.3}_{-2.0}$\\
QCDF &$5.5^{+1.4}_{-2.5}$&$67^{+31}_{-48}$&&\\
SCET &$4.64\pm1.37$&$42\pm14$&&$26.6\pm9.9$\\
FAT&$5.83\pm1.20$&$40.7\pm10.6$&&$29.8\pm5.9$\\
Expt.~\tnote{1}&$4.6\pm1.1$&$78\pm12$&&\\ \hline

$B^0 \to \rho^0 K^{*0}$ &$4.4^{+0.4+0.9+1.9}_{-0.3-0.8-1.3}$&$63.3^{+1.3+10.3+1.3}_{-1.2-9.9-1.5}$
                                &$13.9^{+1.1+4.5+0.7}_{-1.1-4.4-1.0}$&$22.8^{+0.1+5.5+2.4}_{-0.2-5.9-2.0}$\\
PQCD (former) &$3.3^{+1.7}_{-1.4}$&$65^{+4}_{-5}$&&$16.9^{+2.7}_{-1.8}$\\
QCDF &$4.6^{+3.6}_{-3.5}$&$39^{+60}_{-31}$&&\\
SCET &$5.87\pm1.87$&$61\pm13$&&$17.6\pm7.9$\\
FAT&$5.09\pm1.23$&$48.7\pm12.3$&&$25.8\pm6.7$\\
Expt.~\tnote{1}&$3.9\pm1.3$&$40\pm14$&&\\ \hline

$B^0 \to \rho^- K^{*+}$ &$10.5^{+1.2+2.2+4.3}_{-0.9-1.6-3.1}$&$72.9^{+2.2+8.7+2.1}_{-2.1-8.8-3.0}$
                                &$13.6^{+1.1+4.4+1.1}_{-1.2-4.5-1.0}$&$13.5^{+1.0+4.3+1.9}_{-1.1-4.3-1.1}$\\
PQCD (former) &$8.4^{+3.8}_{-3.5}$&$68\pm5$&&$15.6\pm2.5$\\
QCDF &$8.9^{+4.9}_{-5.6}$&$53^{+45}_{-32}$&&\\
SCET &$10.6\pm3.2$&$55\pm14$&&$20.3\pm8.6$\\
FAT&$10.5\pm2.3$&$38.9\pm11.3$&&$30.8\pm6.3$\\
Expt.~\tnote{1}&$10.3\pm2.6$&$38\pm13$&&\\ \hline

$B^0_s\to \rho^+ K^{*-}$ &$34.2^{+12.2+3.4+2.4}_{-8.5-3.3-2.2}$&$91.2^{+0.1+1.0+0.3}_{-0.2-1.3-0.4}$
                                 &$6.8^{+0.1+1.0+0.3}_{-0.0-0.7-0.1}$&$2.0^{+0.1+0.4+0.1}_{-0.1-0.4-0.1}$\\
PQCD (former) &$24.0^{+11.0}_{-9.1}$&$95\pm1$&&$2.31^{+0.22}_{-0.21}$\\
QCDF &$21.6^{+1.6}_{-3.2}$&$92^{+1}_{-4}$&&\\
SCET &$28.1\pm4.2$&$99.1\pm0.3$&&$0.4\pm0.18$\\
FAT&$38.6\pm8.3$&$94.4\pm1.2$&&$2.74\pm0.64$\\
 \hline

$B^0_s\to \rho^0 \bar{K}^{*0}$ &$1.3^{+0.4+0.1+0.3}_{-0.4-0.3-0.4}$&$53.4^{+0.7+6.9+5.4}_{-0.4-6.5-5.2}$
                                       &$25.2^{+0.0+3.3+2.3}_{-0.2-3.4-2.8}$&$21.4^{+0.4+3.3+2.9}_{-0.5-3.5-2.8}$\\
PQCD (former) &$0.40^{+0.22}_{-0.17}$&$57^{+9}_{-13}$&&$22.5^{+7.3}_{-4.7}$\\
QCDF &$1.3^{+2.6}_{-0.7}$&$90^{+5}_{-24}$&&\\
SCET &$1.04\pm0.30$&$87\pm5$&&$5.81\pm2.84$\\
FAT&$1.18\pm0.46$&$79.8\pm8.0$&&$10.2\pm4.1$\\
Expt.~\tnote{1}&$<767$&&& \\
\end{tabular}
\begin{tablenotes}
\item $\tnote{1}$ The experimental results are obtained by multiplying the relevant measured two-body branching
ratios according to the Eq.~(\ref{2body}).
\end{tablenotes}
\end{threeparttable}
\end{ruledtabular}
\end{table}

\subsection{Branching ratios and polarization fractions of two-body $B_{(s)} \to \rho K^*$ decays}
The isospin conservation is assumed for the strong decays of an $I=1/2$ resonance $K^*$ to $K\pi$, namely,
\begin{eqnarray}
\frac{\Gamma(K^{*0} \to K^+\pi^-)}{\Gamma(K^{*0} \to K\pi)}=2/3, ~\quad
\frac{\Gamma(K^{*+} \to K^+\pi^0)}{\Gamma(K^{*+} \to K\pi)}=1/3,
\label{eq:isospin}
\end{eqnarray}
where we assume the $K^*\to K\pi$ branching fraction to be 100\%.
According to the relation of the decay rates between the quasi-two-body and the corresponding two-body decay modes
\begin{eqnarray}\label{2body}
\mathcal{B}(B_{(s)} \rightarrow \rho(\rightarrow \pi\pi)K^*(\rightarrow K\pi))\approx
\mathcal{B}(B_{(s)} \rightarrow \rho K^*)\times \mathcal{B}(\rho \rightarrow \pi\pi)\times \mathcal{B}(K^*\rightarrow K\pi),
\end{eqnarray}
with ${\mathcal B}(\rho \to\pi\pi)=100\%$, we extract the two-body $B_{(s)} \to \rho K^*$ branching ratios and summarize them in Table~\ref{tab:br2body}.
The polarization fractions of the two-body $B_{(s)} \to \rho K^*$ decays calculated in this work are also listed in Table~\ref{tab:br2body}.
For a comparison, we show the updated predictions in the QCDF~\cite{prd80-114026,prd80-114008}, the previous predictions in the PQCD approach~\cite{prd91-054033}, SCET~\cite{prd96-073004} and FAT~\cite{epjc77-333}. Experimental results for branching ratios are taken from Table~\ref{exp1} and for polarization fractions from~\cite{1909-12524}.
\Blue{}
One can see that, except for the colour-suppressed (`` C ") decay  $B^0_s\to \rho^0 \bar{K}^{*}(892)^{0}$, our predictions of the branching ratios are in good agreement with those two-body analyses as presented in the PQCD approach~\cite{prd91-054033}, and also similar to those predicted in the QCDF approach~\cite{prd80-114026,prd80-114008}, SCET~\cite{prd96-073004} and FAT~\cite{epjc77-333} within errors.
However, the situation of the `` C "-type decay $B^0_s\to \rho^0 \bar{K}^{*}(892)^{0}$ is more complicated:
(a) as claimed in Ref.~\cite{plb763-29} that the widths of the resonant states and the interactions between the final state meson pairs will show  their effects on the branching ratios,
the new four-body prediction deviates from the previous calculations in the PQCD approach, but agrees well with the corresponding results in the QCDF approach, SCET and FAT
within errors;
(b) the transverse polarization contribution in the PQCD approach is comparable to the longitudinal one due to the chirally enhanced annihilation and the hard scattering diagrams, which is quite different from those predictions in the QCDF approach, SCET and FAT.
More precise data from the future LHCb and Belle II experiments will help us differentiate these factorization approaches
and understand the underlying mechanism of the multi-body $B$ meson hadronic weak decays.

For the charmless $B_{(s)} \to \rho(\to \pi\pi)K^*(\to K\pi)$ decays, it is naively expected that the helicity amplitudes satisfy the hierarchy pattern $|A_0|\gg|A_+|\gg|A_-|$,
which are related to the spin amplitudes $(A_{0}, A_{\parallel}, A_{\perp})$ in Appendix by
\begin{eqnarray}
\label{eq:helicity}
A_{\pm}=\frac{A_{\parallel}\pm A_{\perp}}{\sqrt{2}},
\end{eqnarray}
while $A_{0}$ is common to both bases.
The above hierarchy relation satisfies the expectation in the factorization assumption that the longitudinal
polarization should dominate based on the quark helicity analysis~\cite{zpc1-269,prd64-117503}.
However, large transverse polarization of order 50$\%$ is observed in $B\to K^*\phi$, $B\to K^*\rho$ and  $B_s\to \phi\phi$ decays, which poses an interesting challenge for the theory.
The interest in the polarization in penguin transition, such as $b\to s$ decays $B \to \rho K^*$, is motivated by their potential sensitivity to physics beyond the SM.
Measurements of the longitudinal polarization fraction in $B \to \rho K^*$ by BABAR~\cite{prl97-201801,prd83-051101,prd85-072005} and Belle~\cite{prl95-141801} reveal a large fraction of transverse polarization, indicating an anomaly of polarization.
An angular analysis of the $B^0 \to \rho^0 K^{*0}$ decay by LHCb measurement found an unexpectedly low longitudinal polarization fraction $f_0=0.164\pm0.015\pm0.022$ where the first uncertainty is statistical and the second systematic~\cite{jhep05-026}.

As shown in Table~\ref{tab:br2body}, the longitudinal polarization fraction $f_0$ for the
$B_{(s)} \to \rho K^*$ decays from the PQCD approach (including the present work) are around 50$\%$ to 80$\%$, which are mostly greater than the transverse one $f_T=f_{\parallel}+f_{\perp}$ in contrast to observations.
The QCDF~\cite{prd80-114008,prd80-114026}, SCET~\cite{prd96-073004} and FAT~\cite{epjc77-333}
yield the similar pattern $f_0\sim f_T$ in despite of large uncertainties.
In the PQCD approach, the large transverse polarization fraction can be interpreted on the basis of the chirally enhanced annihilation diagrams, especially the $(S-P)(S+P)$ penguin annihilation, introduced by the QCD penguin operator $O_6$~\cite{prd71-054025}, which is originally introduced in Ref.~\cite{plb601-151}.
A special feature of the $(S-P)(S+P)$ penguin annihilation operator is that the light quarks in the final states are not produced through chiral currents. So, there is no suppression to the transverse polarization caused by the helicity flip.
Then the polarization fractions satisfy $f_0\approx f_T$.
However, these effects are not able to fully account for the above polarization anomaly.
Our predictions for the longitudinal polarization fractions agree with the previous PQCD calculations~\cite{prd91-054033}.
It is worth mentioning that we have employed the same Gegenbauer moments for
the transversely polarized $K\pi$ DAs as those for the longitudinal polarized ones (see Eq.~(\ref{eq:gen}))
in this work, together with the same Gegenbauer moments for $\pi\pi$ associated with the transverse polarizations from previous work~\cite{prd98-113003}.
The Gegenbauer moments from the twist-3 DAs of $K\pi$ pair may make significant sense to the polarization fractions, which has been verified in Ref.~\cite{plb622-63}.
To be honest, these Gegenbuaer moments should be fitted similarly as those in Ref.~\cite{2105-03899}.
With more and more experimental measurements, we can determine the precise values of these Gegenbauer moments for transversely polarized DAs.

\begin{table}[]
\caption{$CP$-averaged branching ratios (in units of $10^{-6}$) for the four-body $B \to [SS,SV,VS] \to (\pi\pi) (K\pi)$ decays, with $S=[f_0(500),f_0(980),K^*_0(1430)]$ and $V=[\rho, K^*]$. The experimental data are taken from~\cite{pdg2020}. The sources of the theoretical errors are the same as in Table~\ref{tab:br2body}.}
\label{tab:brs}
\begin{ruledtabular}
\begin{threeparttable}
\begin{tabular}{lcc}
Modes &PQCD  & Experiment~\tnote{1}  \\
\hline
$B^+ \to (f_0(500) \to )\pi^+\pi^-(K_0^{*}(1430)^{+}\to)K^0\pi^+$~\tnote{$\text{BW}$} &$17.5^{+8.1+12.0+6.3}_{-5.3-8.1-5.5}$&\\
$B^+ \to (f_0(500) \to )\pi^+\pi^-(K_0^{*}(1430)^{+}\to)K^0\pi^+$~\tnote{$\text{Bugg}$} &$19.2^{+8.9+12.1+7.0}_{-5.9-8.2-6.2}$&\\
$B^0 \to (f_0(500) \to )\pi^+\pi^-(K_0^{*}(1430)^{0}\to)K^+\pi^-$~\tnote{$\text{BW}$} &$18.5^{+8.5+12.2+6.6}_{-5.6-8.3-5.9}$&\\
$B^0 \to (f_0(500) \to )\pi^+\pi^-(K_0^{*}(1430)^{0}\to)K^+\pi^-$~\tnote{$\text{Bugg}$} &$20.3^{+9.4+12.3+7.4}_{-6.2-6.5-9.8}$&\\
$B^0_s\to (f_0(500) \to )\pi^+\pi^-(\bar{K}_0^{*}(1430)^{0}\to)K^-\pi^+$~\tnote{$\text{BW}$} &$0.4^{+0.1+0.6+0.1}_{-0.1-0.3-0.1}$&\\
$B^0_s\to (f_0(500)\to )\pi^+\pi^-(\bar{K}_0^{*}(1430)^{0}\to)K^-\pi^+$~\tnote{$\text{Bugg}$} &$0.4^{+0.1+0.6+0.1}_{-0.1-0.3-0.1}$&\\
$B^+ \to (f_0(980) \to )\pi^+\pi^-(K_0^{*}(1430)^{+}\to)K^0\pi^+$ &$1.6^{+0.4+0.6+0.8}_{-0.3-0.5-0.5}$&\\
$B^0 \to (f_0(980) \to )\pi^+\pi^-(K_0^{*}(1430)^{0}\to)K^+\pi^-$ &$1.5^{+0.4+0.6+0.8}_{-0.3-0.5-0.5}$&$1.2\pm0.4$\\
$B^0_s\to (f_0(980)\to )\pi^+\pi^-(\bar{K}_0^{*}(1430)^{0}\to)K^-\pi^+$ &$0.07^{+0.03+0.03+0.05}_{-0.03-0.03-0.03}$&\\
\hline
$B^+ \to (f_0(500) \to )\pi^+\pi^-(K^{*+}\to)K^0\pi^+$~\tnote{$\text{BW}$} &$1.1^{+0.2+1.5+0.4}_{-0.2-0.2-0.2}$&\\
$B^+ \to (f_0(500) \to )\pi^+\pi^-(K^{*+}\to)K^0\pi^+$~\tnote{$\text{Bugg}$} &$1.1^{+0.2+1.6+0.4}_{-0.2-0.2-0.2}$&\\
$B^0 \to (f_0(500) \to )\pi^+\pi^-(K^{*0}\to)K^+\pi^-$~\tnote{$\text{BW}$}   &$1.0^{+0.2+1.4+0.4}_{-0.2-0.2-0.2}$&\\
$B^0 \to (f_0(500) \to )\pi^+\pi^-(K^{*0}\to)K^+\pi^-$~\tnote{$\text{Bugg}$} &$1.0^{+0.2+1.2+0.4}_{-0.2-0.2-0.2}$&\\
$B^0_s\to (f_0(500) \to )\pi^+\pi^-(\bar{K}^{*0}\to)K^-\pi^+$~\tnote{$\text{BW}$} &$0.17^{+0.04+0.22+0.04}_{-0.04-0.10-0.06}$&\\
$B^0_s\to (f_0(500)\to )\pi^+\pi^-(\bar{K}^{*0}\to)K^-\pi^+$~\tnote{$\text{Bugg}$} &$0.17^{+0.04+0.22+0.04}_{-0.04-0.08-0.04}$&\\
$B^+ \to (f_0(980) \to )\pi^+\pi^-(K^{*+}\to)K^0\pi^+$ &$3.1^{+0.9+0.7+0.7}_{-0.7-0.7-0.9}$&$2.8\pm0.5$\\
$B^0 \to (f_0(980) \to )\pi^+\pi^-(K^{*0}\to)K^+\pi^-$ &$2.9^{+0.9+0.5+1.2}_{-0.7-0.6-0.8}$&$2.6^{+1.4}_{-1.2}$\\
$B^0_s\to (f_0(980)\to )\pi^+\pi^-(\bar{K}^{*0}\to)K^-\pi^+$ &$0.02^{+0.01+0.01+0.01}_{-0.01-0.01-0.01}$&\\
\hline
$B^+ \to (\rho^+ \to )\pi^+\pi^0(K_0^{*}(1430)^{0}\to)K^+\pi^-$ &$14.1^{+6.1+4.9+5.5}_{-3.9-4.3-3.4}$&\\
$B^+ \to (\rho^0 \to )\pi^+\pi^-(K_0^{*}(1430)^{+}\to)K^0\pi^+$ &$5.1^{+2.2+2.1+2.3}_{-1.5-1.8-1.3}$&\\
$B^0 \to (\rho^0 \to )\pi^+\pi^-(K_0^{*}(1430)^{0}\to)K^+\pi^-$ &$7.9^{+3.3+2.3+2.9}_{-2.3-2.2-1.7}$&$18\pm4$\\
$B^0 \to (\rho^- \to )\pi^-\pi^0(K_0^{*}(1430)^{+}\to)K^0\pi^+$ &$11.8^{+5.1+4.2+4.7}_{-3.4-3.7-2.9}$&$19\pm8$\\
$B^0_s\to (\rho^+ \to )\pi^+\pi^0(K_0^{*}(1430)^{-}\to)\bar{K}^0\pi^-$ &$14.1^{+4.3+4.4+1.0}_{-3.2-3.9-1.0}$&\\
$B^0_s\to (\rho^0 \to )\pi^+\pi^-(\bar{K}_0^{*}(1430)^{0}\to)K^-\pi^+$ &$0.4^{+0.1+0.2+0.1}_{-0.1-0.2-0.1}$&
\end{tabular}
\begin{tablenotes}
\item BW denotes the time-like form factor for $f_0(500)$ parameterized by Breit-Wigner formulas.
\item Bugg denotes the time-like form factor for $f_0(500)$ parameterized by D.V.~Bugg model.
\item $\tnote{1}$ The experimental results are obtained by multiplying the relevant measured two-body branching ratios according to the Eq.~(\ref{2body}).
\end{tablenotes}
\end{threeparttable}
\end{ruledtabular}
\end{table}
\subsection{Branching ratios of $B_{(s)} \to [SS,SV,VS]\to (\pi\pi)(K\pi)$ decays}
In contrast to the vector resonances, the identification of the scalar mesons is a long-standing puzzle.
Scalar resonances are difficult to resolve because some of them have large decay widths, which cause a strong overlap between resonances and background.
In fact, compared with the $B_{(s)} \to VV \to (\pi\pi)(K\pi)$ decays, there are much less experimental data for the $B_{(s)} \to [SS,SV,VS]\to (\pi\pi)(K\pi)$ decays.
Furthermore, the underlying structure of scalar mesons is not theoretically well established (for a review, see Ref.~\cite{pdg2020}).
We hope that the situation can be improved using nonperturbative QCD tools including lattice QCD simulations.
The $f_0(980)$ is strongly produced in $D^+_s$ decay~\cite{prl86-765}, which implies a large $\bar{s}s$ component, assuming Cabibbo-favored $c \to s$ decay.
Meanwhile, the prominent appearance of the $f_0(980)$ points to a dominant ($\bar{s}s$) component in the semileptonic $D_s$ decays and decays of $B_{(s)}$ mesons.
However, there also exists some experimental evidences indicating that $f_0(980)$ is not purely an $\bar{s}s$ state.
Ratios of decay rates of $B$ and/or $B_s$ mesons into $J/\psi$
plus $f_0(980)$ or $f_0(500)$ were proposed to allow for an extraction of
the flavor mixing angle and to probe the tetraquark nature of those
mesons within a certain model~\cite{epjc71-1832,prl111-062001}.
The phenomenological fits of the LHCb do neither support a contribution of the $f_0(980)$ in the $B\to J/\psi\pi\pi$~\cite{prd90-012003} nor an $f_0(500)$ in $B_s \to J/\psi\pi\pi$ decays~\cite{prd89-092006} by employing the isobar model.
Hence the authors conclude that their data is incompatible with a model where $f_0(980)$ is formed from two quarks and two antiquarks (tetraquarks) at the eight standard deviation level.
In addition, they extract an upper limit for the mixing angle of $17^\circ$ at $90\%$ confidence level between
the $f_0(980)$ and the $f_0(500)$ that would yield a substantial
($\bar{s}s$) content in $f_0(980)$~\cite{prd90-012003}.
But in fact a substantial $f_0(980)$ contribution is also found in the $B$-decays in a dispersive analysis of the same data that allows for a model-independent inclusion of the hadronic final state interactions in Ref.~\cite{jhep1602-009}, which puts into question the
conclusions of Ref.~\cite{prd90-012003}.
At this stage, the quark structure of scalar particles are still quite controversial.
On the theory side, there are some studies on the $f_0(980)$ by assuming the $f_0(980)$ as a pure $\bar{s}s$ state.
For example, the authors studied the $B_s \to J/\psi f_0(980)$ with the light-cone QCD sum rule and factorization assumption in Ref.~\cite{prd81-074001} and using generalized factorization
and SU(3) flavor symmetry in Ref.~\cite{prd83-094027}.
In Ref.~\cite{epjc80-554}, the authors calculated the $\bar{B}_s \to f_0(980)$ form factor from the light-cone sum rules with $B$-meson DAs, and
investigated the $S$-wave $\bar{B}_s \to KK$ form factors to study the width effect, where the $f_0(980)$ is dominated by the $\bar{s}s$ configuration.
As a first approximation, we take into account the scalar meson $f_0(500),f_0(980),K_0^*(1430)$ in the $\bar{q}{q}$ density operator with $q=(u,d,s)$.
The $S$-wave time-like form factor $F_S(\omega^2)$ used to parameterize the $S$-wave two-pion and kaon-pion DAs have been determined in Refs.~\cite{prd91-094024,epjc76-675}.

We list the branching ratios of the four-body decays $B_{(s)} \to [SS,SV,VS]\to (\pi\pi)(K\pi)$ with experimental data~\cite{pdg2020} in Table~\ref{tab:brs}.
So far, only five of them, say $B^{+} \to (f_0(980) \to )\pi^+\pi^-(K^{*+}\to)K^{0}\pi^{+}$,
$B^{0} \to (f_0(980) \to )\pi^+\pi^-(K^{*0}\to)K^{+}\pi^{-}$,
$B^{0} \to (\rho^{-} \to )\pi^{-}\pi^{0}(K_0^{*}(1430)^{+}\to)K^{0}\pi^{+}$,
$B^{0} \to (\rho^{0} \to )\pi^{+}\pi^{-}(K_0^{*}(1430)^{0}\to)K^{+}\pi^{-}$,
and $B^0 \to (f_0(980) \to )\pi^+\pi^-(K_0^{*}(1430)^{0}\to)K^+\pi^-$, have been reported by experiments.
It is shown that, except for the colour suppressed (`` C ") decay
$B^{0} \to (\rho^{0} \to )\pi^{+}\pi^{-}(K_0^{*}(1430)^{0}\to)K^{+}\pi^{-}$,
our predictions of other four channels are consistent with the available experimental data within errors,
with the remaining predictions awaiting for the examinations from future experimental measurements.
However, the branching ratio ${\cal B}(B^{0} \to (\rho^{0} \to )\pi^{+}\pi^{-}(K_0^{*}(1430)^{0}\to)K^{+}\pi^{-})=(7.9^{+4.9}_{-3.6})\times 10^{-6}$ estimated in this work is smaller than the experimental data ${\cal B}=(18\pm 4)\times 10^{-6}$~\cite{pdg2020} by a factor of $\sim 2$.
Since only leading order contributions are considered in this work, it indicates that this decay mode might be more sensitive to next-to-leading order corrections,
and it is similar to the situation of other `` C "-type decays, such as $B^0\to \pi^0\pi^0,\rho^0\rho^0$.
Besides, under the isospin limit, it is naively expected that
\begin{eqnarray} \label{eq:ratios}
 R=\frac{{\cal B}(B^0 \to (\rho^0 \to )\pi^+\pi^-(K_0^{*}(1430)^{0}\to)K^+\pi^-)}{{\cal B}(B^0 \to (\rho^- \to )\pi^-\pi^0(K_0^{*}(1430)^{+}\to)K^0\pi^+)}=\frac{1}{2},
\end{eqnarray}
which is not borne out by experiment and needs to be further studied in the future.
Among the three different kinds of theoretical errors considered in our work, one can see that the most important theoretical uncertainties for the branching ratio are caused by the nonperturbative input parameters of the wave functions for some decay modes.
Taking the decay $B^+ \to (f_0(500) \to )\pi^+\pi^-(K^{*+}\to)K^0\pi^+$ as an example,
which is dominated by the $B \to (f_0(500) \to) \pi\pi$ transition progress,
its branching ratio is much more sensitive to the Gegenbauer moment $a_S$ from the $S$-wave DAs.
In the PQCD approach, wave functions are the most important input parameters and the improved knowledge of them is expected to yield improved estimates of the branching ratios and other observables, which may lead to better consistency with the data.

Since the $f_0(500)$ is very broad, we use BW formula and Bugg model to parameterize the $f_0(500)$ resonance and compare their results.
It is found that the model-dependence of the decay rate is indeed not significant.
The central values of PQCD predictions based on the Bugg model are consistent with the ones from the BW formula.
Our prediction of $B^0\to (f_0(980)\to)\pi^+\pi^-(K^{*0}\to)K^+\pi^-$ is consistent with the current data, and also comparable with that from Table~III in~\cite{2008-08458} within errors.
In order to compare our predictions with other theoretical results for decays involving $f_0(980)$, we use the ${\cal B}(f_0(980)\to \pi^+\pi^-)=0.50$, which is taken from~\cite{prd87-114001} and in agreement with the value of ${\cal B}(f_0(980)\to \pi^+\pi^-)=0.46$ obtained in~\cite{prd87-052001}.
We can extract the branching ratios of the two-body decays $B \to f_0(980)K^*$ from the corresponding four-body decays $B \to f_0(980)(\to \pi\pi)K^*(\to K\pi)$ in Table~\ref{tab:brs} under the narrow width approximation.
Taking the decay $B^0\to f_0(980)K^{*0}$ as an example, we obtain its branching ratio ${\cal B}(B^0\to f_0(980)K^{*0})=8.7\times10^{-6}$, which is in good agreement with previous two-body results in the QCDF approach~\cite{prd87-114001} and PQCD approach~\cite{epjc67-163}.
Strictly speaking, the narrow width approximation is not fully justified since such approximation has its scope of application.
As mentioned above, the nonperturbative input parameters from the wave functions make important sense to the branching ratios.
We can fit the related Gegenbauer moments with abundant data to match the experiment in the future.
However, the fact that their rates can be accommodated in the two-quark picture for $f_0(980)$ does not mean that $\bar{q}{q}$ composition should be supported.
It is too difficult to make theoretical predictions on these decay modes based on the four-quark picture for scalar resonances.
We just assume they are constituted by two quarks at this moment.

The decays $B \to \rho K_0^{*}(1430)$ have already been studied systematically in the two-body framework within the PQCD approach~\cite{prd82-034036}.
Taking the two measured channels $B^{0} \to \rho^{-}K_0^{*}(1430)^{+} \to (\pi^{-}\pi^{0})(K^{0}\pi^{+})$ and
$B^{0} \to \rho^{0}K_0^{*}(1430)^{0} \to (\pi^{+}\pi^{-})(K^{+}\pi^{-})$ as examples,
we have:
\begin{eqnarray}
{\cal B }(B^0 \to \rho^- K_0^{*}(1430)^{+} \to (\pi^-\pi^0)(K^0\pi^+))=
\left\{\begin{array}{ll}
(11.8^{+8.1}_{-5.8})\times 10^{-6}                & {\rm this \quad work},\\
(7.0^{+2.9}_{-1.7})\times 10^{-6}         & {\rm PQCD}~[16],\\
(19\pm 8)\times 10^{-6}                 & {\rm Data} ~[46],\\
\end{array} \right.
\end{eqnarray}
\begin{eqnarray}
{\cal B }(B^0 \to \rho^0 K_0^{*}(1430)^{0}\to (\pi^+\pi^-)(K^+\pi^-))=
\left\{\begin{array}{ll}
(7.9^{+4.9}_{-3.6})\times 10^{-6}         & {\rm this \quad work},\\
(3.2^{+1.0}_{-0.7})\times 10^{-6}         & {\rm PQCD}~[16],\\
(18\pm 4)\times 10^{-6}                  & {\rm Data} ~[46].\\
\end{array} \right.
\end{eqnarray}
The results from the previous PQCD work~\cite{prd82-034036} are obtained by multiplying the relevant two-body branching ratios according to Eqs.~(\ref{eq:isospin})-(\ref{2body}).
Since the width of the resonant state and the interactions between the final state meson pair will show their effects on the branching ratios, the new four-body predictions are relatively larger than the converted values from previous PQCD calculations, but more close to the experimental data.
Therefore, it seems more appropriate to treat these decay modes as four-body decays.

\begin{table}[t]
\caption{Direct $CP$ asymmetries (in units of $\%$) for the $B_{(s)} \to VV \to (\pi\pi)(K\pi)$ decays compared with the previous predictions in the PQCD approach~\cite{prd91-054033}, the updated predictions in the QCDF~\cite{prd80-114008,prd80-114026},  SCET~\cite{prd96-073004} and FAT~\cite{epjc77-333}.
Experimental results for branching ratios are taken from Table~\ref{exp2}.
The sources of the theoretical errors are the same as in Table~\ref{tab:br2body}.}
\label{tab:cpv}
\begin{ruledtabular}

\begin{tabular}[t]{lcccc}
Modes  & $\mathcal{A}_0^{\text{CP}}$ & $\mathcal{A}_{\parallel}^{\text{CP}}$& $\mathcal{A}_{\perp}^{\text{CP}}$ & $\mathcal{A}^{\text{CP}}$\\
\hline
$B^+ \to (\rho^+ \to )\pi^+\pi^0(K^{*0}\to)K^+\pi^-$ &$0.3^{+0.3+0.7+2.2}_{-0.1-0.1-1.1}$&$1.2^{+0.0+0.5+1.0}_{-0.1-1.3-1.7}$
                                                     &$2.5^{+0.0+0.9+0.5}_{-0.4-1.1-1.8}$&$0.7^{+0.2+0.6+1.7}_{-0.1-0.3-1.1}$\\
PQCD (former) &$-0.36^{+0.12}_{-0.11}$&&$0.98^{+0.20}_{-0.25}$&$-1.0^{+0.3}_{-0.4}$\\
QCDF &&&&$-0.3^{+2}_{-0}$\\
SCET &$-0.68\pm0.77$&&$0.56\pm0.61$&$-0.56\pm0.61$\\
FAT&$1.40\pm0.56$&&$-1.19\pm0.19$&$1.00\pm0.17$\\
Expt. &&&&$-1\pm16$\\ \hline

$B^+ \to (\rho^0 \to )\pi^+\pi^-(K^{*+}\to)K^0\pi^+$ &$20.5^{+0.3+2.3+6.0}_{-0.2-2.5-5.2}$&$1.1^{+0.1+4.6+1.3}_{-0.8-4.6-1.5}$
                                                     &$-53.8^{+5.3+5.5+7.4}_{-5.8-6.4-6.7}$&$11.8^{+0.3+3.8+3.8}_{-0.3-4.3-3.7}$\\
PQCD (former) &$11.3^{+2.3}_{-2.4}$&&$-34.0^{+3.7}_{-2.8}$&$22.7^{+2.9}_{-3.2}$\\
QCDF&&&&$43^{+13.4}_{-28}$\\
SCET &$40.4\pm51.3$&&$-29.3\pm31.0$&$29.3\pm{31.0}$\\
FAT&$35.0\pm19.8$&&$-24.2\pm9.0$&$34.6\pm8.3$\\
Expt. &&&&$31\pm13$ \\\hline

$B^0 \to (\rho^0 \to )\pi^+\pi^-(K^{*0}\to)K^+\pi^-$ &$3.5^{+0.8+4.1+6.0}_{-0.6-2.5-4.5}$&$-35.9^{+4.8+6.6+5.1}_{-4.9-9.6-5.7}$
                                                     &$12.3^{+0.0+3.2+2.9}_{-0.4-3.2-3.1}$&$0.04^{+0.0+2.3+3.6}_{-0.1-2.6-3.0}$\\
PQCD (former) &$3.64^{+1.20}_{-1.07}$&&$-7.71^{+1.97}_{-1.86}$&$-8.9^{+3.1}_{-3.0}$\\
QCDF &&&&$-15^{+16.5}_{-16.1}$\\
SCET &$-2.10\pm2.67$&&$3.30\pm3.91$&$-3.30\pm3.91$\\
FAT&$-0.41\pm4.3$&&$0.39\pm4.06$&$-0.6\pm4.0$\\
Expt. &&&&$-6\pm9$ \\\hline

$B^0 \to (\rho^- \to )\pi^-\pi^0(K^{*+}\to)K^0\pi^+$ &$25.0^{+0.9+2.2+8.5}_{-0.5-2.9-5.9}$&$-23.2^{+2.6+3.8+3.1}_{-3.1-4.3-3.1}$
                                                     &$-27.0^{+2.7+2.7+3.6}_{-3.1-2.7-3.1}$&$11.5^{+1.0+4.7+4.6}_{-0.7-5.2-3.8}$\\
PQCD (former) &$23.8^{+4.7}_{-5.1}$&&$-50.9^{+4.9}_{-3.9}$&$24.5^{+3.1}_{-3.8}$\\
QCDF &&&&$32^{+2.2}_{-14.3}$\\
SCET &$16.8\pm21.7$&&$-20.6\pm23.3$&$20.6\pm23.3$\\
FAT&$37.2\pm18.9$&&$-23.8\pm6.9$&$34.3\pm6.3$\\
Expt. &&&&$21\pm15$\\ \hline

$B^0_s\to (\rho^+ \to )\pi^+\pi^0(K^{*-}\to){\bar K}^0\pi^-$ &$-13.6^{+1.8+1.5+1.9}_{-1.7-1.3-2.0}$&$30.8^{+4.6+6.3+6.0}_{-4.2-6.5-5.1}$
                                                      &$54.3^{+6.6+10.8+9.0}_{-6.5-11.4-9.2}$&$-9.2^{+1.2+1.4+1.5}_{-1.0-1.1-1.2}$\\
PQCD (former) &$-2.71^{+0.68}_{-0.72}$&&$55.0^{+10.3}_{-10.5}$&$-9.1^{+1.7}_{-1.9}$\\
QCDF &&&&$-11^{+4.1}_{-1.4}$\\
SCET &$-0.07\pm0.09$&&$7.68\pm9.19$&$-7.68\pm9.19$\\
FAT&$0.91\pm0.45$&&$-15.4\pm9.5$&$-10.9\pm3.0$\\ \hline

$B^0_s\to (\rho^0 \to )\pi^+\pi^-(\bar{K}^{*0}\to)K^-\pi^+$ &$13.8^{+1.3+6.9+18.7}_{-3.5-11.5-17.6}$&$34.9^{+4.8+13.2+7.0}_{-5.0-13.5-6.0}$
                                                            &$41.9^{+4.7+13.1+5.3}_{-5.0-13.3-8.9}$&$25.1^{+0.5+6.4+7.3}_{-1.7-8.2-9.2}$\\
PQCD (former) &$-17.5^{+21.2}_{-13.0}$&&$22.0^{+29.9}_{-31.4}$&$62.7^{+14.4}_{-18.8}$\\
QCDF &&&&$46^{+18}_{-30}$\\
SCET &$2.87\pm4.00$&&$-19.5\pm23.5$&$19.5\pm23.5$\\
FAT&$0.47\pm4.69$&&$-1.89\pm18.3$&$4.9\pm18.3$\\
\end{tabular}
\end{ruledtabular}
\end{table}
\begin{table}[]
\caption{Direct $CP$ asymmetries (in units of $\%$) for the four-body $B \to [SS,SV,VS] \to (\pi\pi) (K\pi)$ decays, with $S=[f_0(500),f_0(980),K^*_0(1430)]$ and $V=[\rho, K^*]$. The sources of the theoretical errors are the same as in Table~\ref{tab:br2body}.}
\label{tab:cps}
\begin{ruledtabular}
\begin{threeparttable}
\begin{tabular}{lc}
Modes &PQCD   \\
\hline
$B^+ \to (f_0(500) \to )\pi^+\pi^-(K_0^{*}(1430)^{+}\to)K^0\pi^+$~\tnote{$\text{BW}$} &$4.1^{+0.8+1.9+0.6}_{-0.7-2.6-0.4}$\\
$B^+ \to (f_0(500) \to )\pi^+\pi^-(K_0^{*}(1430)^{+}\to)K^0\pi^+$~\tnote{$\text{Bugg}$} &$4.1^{+0.9+1.7+0.5}_{-0.7-2.9-0.4}$\\
$B^0 \to (f_0(500) \to )\pi^+\pi^-(K_0^{*}(1430)^{0}\to)K^+\pi^-$~\tnote{$\text{BW}$} &$3.4^{+0.6+2.1+0.9}_{-0.5-3.2-0.2}$\\
$B^0 \to (f_0(500) \to )\pi^+\pi^-(K_0^{*}(1430)^{0}\to)K^+\pi^-$~\tnote{$\text{Bugg}$} &$3.4^{+0.6+2.0+0.6}_{-0.5-3.2-0.2}$\\
$B^0_s\to (f_0(500) \to )\pi^+\pi^-(\bar{K}_0^{*}(1430)^{0}\to)K^-\pi^+$~\tnote{$\text{BW}$} &$-77.0^{+7.9+22.1+13.9}_{-6.8-4.1-5.3}$\\
$B^0_s\to (f_0(500)\to )\pi^+\pi^-(\bar{K}_0^{*}(1430)^{0}\to)K^-\pi^+$~\tnote{$\text{Bugg}$} &$-76.9^{+8.5+20.8+13.7}_{-6.8-4.3-5.3}$\\
$B^+ \to (f_0(980) \to )\pi^+\pi^-(K_0^{*}(1430)^{+}\to)K^0\pi^+$ &$-0.3^{+0.8+1.2+1.1}_{-0.0-1.2-2.7}$\\
$B^0 \to (f_0(980) \to )\pi^+\pi^-(K_0^{*}(1430)^{0}\to)K^+\pi^-$ &$-0.3^{+0.3+1.1+3.3}_{-0.0-0.9-2.3}$\\
$B^0_s\to (f_0(980)\to )\pi^+\pi^-(\bar{K}_0^{*}(1430)^{0}\to)K^-\pi^+$ &$2.8^{+0.8+0.7+1.9}_{-0.6-1.5-1.6}$\\
\hline
$B^+ \to (f_0(500) \to )\pi^+\pi^-(K^{*+}\to)K^0\pi^+$~\tnote{$\text{BW}$} &$-31.4^{+3.5+19.3+4.7}_{-4.5-4.7-7.6}$\\
$B^+ \to (f_0(500) \to )\pi^+\pi^-(K^{*+}\to)K^0\pi^+$~\tnote{$\text{Bugg}$} &$-34.4^{+3.5+19.6+3.1}_{-4.5-5.4-7.2}$\\
$B^0 \to (f_0(500) \to )\pi^+\pi^-(K^{*0}\to)K^+\pi^-$~\tnote{$\text{BW}$} &$17.9^{+0.5+0.6+2.4}_{-1.4-17.3-6.4}$\\
$B^0 \to (f_0(500) \to )\pi^+\pi^-(K^{*0}\to)K^+\pi^-$~\tnote{$\text{Bugg}$} &$18.3^{+0.1+0.6+1.7}_{-0.3-18.0-6.3}$\\
$B^0_s\to (f_0(500) \to )\pi^+\pi^-(\bar{K}^{*0}\to)K^-\pi^+$~\tnote{$\text{BW}$} &$41.3^{+1.0+2.9+6.3}_{-1.2-41.4-7.5}$\\
$B^0_s\to (f_0(500)\to )\pi^+\pi^-(\bar{K}^{*0}\to)K^-\pi^+$~\tnote{$\text{Bugg}$} &$39.1^{+1.1+4.2+7.5}_{-0.5-39.6-7.2}$\\
$B^+ \to (f_0(980) \to )\pi^+\pi^-(K^{*+}\to)K^0\pi^+$ &$0.2^{+0.3+0.7+0.5}_{-0.4-0.9-0.4}$\\
$B^0 \to (f_0(980) \to )\pi^+\pi^-(K^{*0}\to)K^+\pi^-$ &$-0.2^{+0.0+0.9+1.4}_{-0.0-0.6-0.4}$\\
$B^0_s\to (f_0(980)\to )\pi^+\pi^-(\bar{K}^{*0}\to)K^-\pi^+$ &$1.3^{+0.2+0.8+0.1}_{-0.4-0.1-1.8}$\\
\hline
$B^+ \to (\rho^+ \to )\pi^+\pi^0(K_0^{*}(1430)^{0}\to)K^+\pi^-$ &$2.5^{+1.0+1.2+1.2}_{-0.4-1.2-1.2}$\\
$B^+ \to (\rho^0 \to )\pi^+\pi^-(K_0^{*}(1430)^{+}\to)K^0\pi^+$ &$-3.6^{+0.0+1.6+0.0}_{-1.4-7.3-4.8}$\\
$B^0 \to (\rho^0 \to )\pi^+\pi^-(K_0^{*}(1430)^{0}\to)K^+\pi^-$ &$6.8^{+1.5+4.7+0.0}_{-1.3-3.9-1.6}$\\
$B^0 \to (\rho^- \to )\pi^-\pi^0(K_0^{*}(1430)^{+}\to)K^0\pi^+$ &$2.4^{+0.1+1.0+0.0}_{-1.4-3.8-3.4}$\\
$B^0_s\to (\rho^+ \to )\pi^+\pi^0(K_0^{*}(1430)^{-}\to)\bar{K}^0\pi^-$ &$7.8^{+1.1+0.7+1.6}_{-1.0-1.0-1.3}$\\
$B^0_s\to (\rho^0 \to )\pi^+\pi^-(\bar{K}_0^{*}(1430)^{0}\to)K^-\pi^+$ &$56.3^{+4.0+10.4+12.1}_{-3.1-9.7-5.6}$
\end{tabular}
\begin{tablenotes}
\item BW denotes the time-like form factor for $f_0(500)$ parameterized by Breit-Wigner formulas.
\item Bugg denotes the time-like form factor for $f_0(500)$ parameterized by D.V.~Bugg model.
\end{tablenotes}
\end{threeparttable}
\end{ruledtabular}
\end{table}
\subsection{Direct $CP$ asymmetries}
In Table~\ref{tab:cpv}, we show the direct $CP$ asymmetries with each helicity state ($\mathcal{A}_{0,\parallel,\perp}^{\text{CP}}$) for the four-body $B_{(s)} \to VV \to (\pi\pi) (K\pi)$ decays together with those summed over all helicity states ($\mathcal{A}^{\text{CP}}$).
For comparison, the updated results of the QCDF~\cite{prd80-114008,prd80-114026}, SCET~\cite{prd96-073004} and FAT~\cite{epjc77-333} as well as the PQCD predictions in two-body framework~\cite{prd91-054033} are also presented.
Meanwhile, direct $CP$ asymmetries for the four-body $B_{(s)} \to [SS,SV,VS] \to (\pi\pi) (K\pi)$ decays are displayed in Table~\ref{tab:cps}.
As we know, the kinematics of the two-body decays is fixed, the decay amplitudes of the quasi-two-body decays depend on the invariant mass of the final-state pairs,
which result in the differential distribution of direct $CP$ asymmetries.
The $CP$ asymmetry in the four-body framework is moderated by the finite width of the intermediate resonance appearing in the time-like form factor $F(\omega^2)$.
Thus, it is reasonable to see the differences of direct $CP$ asymmetries between the two-body and four-body frameworks in the PQCD approach.
By comparing the numerical results as listed in Table~\ref{tab:cpv}, due to the different mechanism and origins of the strong phase,
one can see that the QCDF and SCET results for the direct $CP$ asymmetries are quite different from ours for some decay modes.
As is well known, besides the weak phase from the CKM matrix elements, the direct $CP$ asymmetry is proportional to the strong phase.
In the SCET, the strong phase is only from the nonperturbative charming penguin at leading power and leading order, while in the QCDF and PQCD approaches,
the strong phase comes from the hard spectator scattering and annihilation diagrams respectively.
Besides, the power corrections such as penguin annihilation, which are essential to resolve the $CP$ puzzles in the QCDF,
are often plagued by the endpoint divergence that in turn break the factorization theorem~\cite{prd80-114008}.
In the PQCD approach, the endpoint singularity is cured by including the parton's transverse momentum.
Anyway, since current experimental measurements still have relatively large uncertainties, we have to wait for more time to test these different predictions.

In Tables~\ref{tab:cpv} and~\ref{tab:cps}, a large $CP$ asymmetry can be understood due to the sizable interference between the tree and penguin amplitudes, while a small value of $CP$ asymmetry is attributed to the dominant tree or penguin amplitudes.
For example, among the six considered $B_{(s)}\to \rho K^*\to(\pi\pi)(\pi K)$ decays as presented in Table~\ref{tab:cpv}, the $CP$ asymmetries $\mathcal{A}^{\text{CP}}$ for the two penguin-dominant processes $B^0 \to (\rho^0\to)\pi^+\pi^- (K^{*0}\to)K^+\pi^-$ and $B^+ \to (\rho^+\to)\pi^+\pi^0 (K^{*0}\to)K^+\pi^-$ are indeed quite small: less than $1\%$.
However, for the `` Color-suppressed " decay $B_s^0 \to (\rho^0\to)\pi^+\pi^- ({\bar K}^{*0}\to)K^-\pi^+$, due to the large penguin contributions from the chirally enhanced annihilation diagrams,
the sizable interference between the tree and penguin contributions makes the direct $CP$ asymmetries $\mathcal{A}^{\text{CP}}$ as large as $\sim 30\%$.
For four $B^{+,0} \to \rho K^* \to (\pi\pi)(K\pi)$ decays, our predictions of $CP$ asymmetries are in agreement with observations within uncertainties.
Moreover, a helicity-specific analysis would provide interesting further insights.
Very recently, LHCb~\cite{jhep05-026} has reported the $CP$ asymmetry associated with longitudinal polarization $\mathcal{A}^0_{\rho K^*}=-0.62\pm0.09\pm0.09$, where the first uncertainty is statistical and the second systematic.
The data is much different from our prediction $\mathcal{A}_0^{\text{CP}}(B^0 \to \rho^0(\to \pi^+\pi^-)K^{*0}(\to K^+\pi^-))=3.5\%$.
Considering the branching ratio of the $B^0 \to \rho^0(\to \pi^+\pi^-)K^{*0}(\to K^+\pi^-)$ decay associated with the longitudinal polarization, the contributions of the penguin diagrams (${\cal B}=2.98\times 10^{-6}$) are larger than the tree ones (${\cal B}=5.74 \times 10^{-8}$) by roughly a factor of 52, which results in the smallness of direct $CP$ asymmetries.
The big gap between the theory and experiment should be resolved in the future.

In the limit of $U$-spin symmetry, some of $B_s$ decays can be related to $B^0$ ones.
For $B_{(s)} \to VV$ decays, it has been  studied in~\cite{prd80-114026,prd91-054033} and  seems to hold well.
Since we have calculated the $B$ and $B_s$ decays to $VV$ in this work in the PQCD approach, we also check the $U$-spin symmetry in some decay modes studied in \cite{prd80-114026,prd91-054033}:
\begin{eqnarray}
&&\mathcal{A}^{\text{CP}}(B_s\to \rho^+K^{*-})=-\mathcal{A}^{\text{CP}}(B^0\to \rho^-K^{*+})\frac{\mathcal{B}(B^0\to\rho^-K^{*+})}{\mathcal{B}(B_s\to\rho^+K^{*-})}\frac{\tau_{B_{s}}}{\tau_{B^0}},\nonumber\\
&&\mathcal{A}^{\text{CP}}(B_s\to \rho^0\bar{K}^{*0})=-\mathcal{A}^{\text{CP}}(B^0\to \rho^0 K^{*0})\frac{\mathcal{B}(B^0\to\rho^0 K^{*0})}{\mathcal{B}(B_s\to \rho^0\bar{K}^{*0})}\frac{\tau_{B_{s}}}{\tau_{B^0}}.
\end{eqnarray}
On basis of these $U$-spin relations along with the branching ratios, the lifetimes of $B$ and $B_s$ mesons and the direct $CP$ asymmetries in $B$ decays, we can get the relevant direct $CP$ asymmetries in $B_s$ decays.
This can be then compared with the corresponding predictions in the PQCD approach to check whether the $U$-spin symmetry works well or not.
We show this comparison in Table~\ref{tb:u-spin}, where the entries in the last two columns have to be compared with each other.
It turns out that $U$-spin symmetry is in general acceptable within the calculational errors.
\begin{table*}[h]
 \caption{The direct $CP$ asymmetries ($\%$) in $B_s \to \rho K^*$ decays via $U$-spin relation together with the direct PQCD prediction. }
\begin{ruledtabular}
 \begin{tabular}{lcclccc}
 Modes & ${\cal B} (10^{-6})$ &$\mathcal{A}^{\text{CP}}$($\%$)~~&Modes&${\cal B} (10^{-6})$&$\mathcal{A}^{\text{CP}}$($\%$)($U$)&$\mathcal{A}^{\text{CP}}$(PQCD) \\
 \hline
$B^0\to \rho^-K^{*+}$&10.5&$11.5^{+6.7}_{-6.5}$&$B_s\to \rho^+K^{*-}$&34.2&-3.5&$-9.2^{+2.4}_{-1.9}$\\
$B^0\to \rho^0 K^{*0}$&4.4&$0.04^{+4.3}_{-4.0}$&$B_s\to \rho^0\bar{K}^{*0}$&1.3&-0.13&$25.1^{+9.7}_{-12.4}$\\
\end{tabular} \label{tb:u-spin}
\end{ruledtabular}
\end{table*}

\subsection{Triple  product asymmetries in $B_{(s)} \to \rho(\to \pi\pi)K^*(\to K\pi)$ decays}
The predicted TPAs  for the  $B_{(s)} \to (\pi\pi)(K\pi)$  decays are displayed in Table~\ref{tab:tpas}.
It is shown that our PQCD predictions of ``true" $CP$-violating TPAs are very small in the SM, which makes the measurement of a large value for that TPA point clearly towards the presence of new physics.
As ``fake'' TPAs are due to strong phases and require no $CP$ violation, the large fake $\mathcal{A}_{\text{T-fake}}^{1,2}$ simply reflects the importance of the strong final-state phases.

Since the left-handedness of the weak interaction $A_{-}\ll A_{+}$ is expected, it implies $A_{\parallel}\approx A_{\perp}$.
The $\mathcal{A}_T^2$ term requires both transversely polarized components $A_{\parallel}$ and $A_{\perp}$ and the decay amplitude associated with transverse polarization is smaller than that for longitudinal polarization in the naive expectation.
Hence $\mathcal{A}_T^2$ is power suppressed relative to $\mathcal{A}_T^1$.
Meanwhile, the smallness of $\mathcal{A}_T^2$  is also attributed to the suppression from the strong phase
difference between the perpendicular and parallel polarization amplitudes, which was found in the PQCD framework~\cite{prd91-054033} and supported by the LHCb Collaboration~\cite{jhep05-026}.
An observation of $\mathcal{A}_T^2$ with large values can signify physics beyond the SM.
As mentioned above, $(\mathcal{A}_T^{1,2}+\bar{\mathcal{A}}_T^{1,2})/2\neq \mathcal{A}_T^{(1,2)\text{ave}}(\text{true})$ when the decay channel has a nonzero $CP$ asymmetries.
We find that the greater difference between the $\mathcal{A}_T^{1,2}(\text{true})$ and $\mathcal{A}_T^{(1,2)\text{ave}}(\text{true})$ appears with the larger direct $CP$ asymmetry.

Recently, the measurements of ``true" and ``fake" TPAs for $B^0 \to \rho^0 K^{*0} \to (\pi^+ \pi^-)(K^+\pi^-)$ have been reported by LHCb Collaboration~\cite{jhep05-026}.
The PQCD prediction of  $\mathcal{A}_{\text{T-true}}^1$ agrees well with the experiment $\mathcal{A}_{\text{T-true}}^{\rho K^*,1}=-0.0210\pm0.0050\pm0.0022$, where the first uncertainty is statistical and the second systematic.
While for ``fake" TPAs, our predictions are a little larger than the measurements but compatible within large uncertainties.
It should be stressed that there are large uncertainties in both experimental measurements and the theoretical calculations for TPAs, so the discrepancy between the data and the theoretical results could be clarified with the high precision both in experimental and theoretical sides.
Since ``fake" TPAs strongly affected by the strong phases,  we lack a perfect knowledge of all the possible signals of the strong phases, such as final-state interactions.
For this reason, we just estimate the size of the corresponding TPAs.
We hope the future experiments can test our predictions.

\begin{table}[t]
\caption{PQCD predictions for the TPAs ($\%$) of the four-body $B_{(s)} \to (\rho \to) \pi\pi (K^*\to) K\pi$ decays.
The sources of theoretical errors are same as in Table~\ref{tab:br2body} but added in quadrature.}
\label{tab:tpas}
\begin{center}
\begin{threeparttable}
\begin{tabular}{l|c|c|c|c|c|c}
\hline\hline
\multicolumn{1}{c|}{}  &\multicolumn{6}{c}{ $\text{TPAs}$-1}   \cr\cline{2-7}
{Modes} &$\mathcal{A}_{T}^1$&$\bar{\mathcal{A}}_{T}^1$&$\mathcal{A}_{\text{T-true}}^1$ &$\mathcal{A}_{\text{T-fake}}^1$&$\mathcal{A}_{\text{T-True}}^{(1)\text{ave}}$ &$\mathcal{A}_{\text{T-fake}}^{(1)\text{ave}}$  \cr \hline
$B^+ \to (\rho^+ \to )\pi^+\pi^0(K^{*0}\to)K^+\pi^-$ &$24.94^{+2.05}_{-3.27}$&$-25.65^{+3.81}_{-2.72}$&$-0.36^{+0.28}_{-0.54}$
                                             &$25.29^{+2.37}_{-3.33}$&$-0.07^{+0.29}_{-0.56}$&$25.29^{+2.36}_{-3.34}$\\ \hline

$B^+ \to (\rho^0 \to )\pi^+\pi^-(K^{*+}\to)K^0\pi^+$ &$14.51^{+3.55}_{-4.06}$&$-24.52^{+3.37}_{-2.35}$&$-5.00^{+1.44}_{-1.33}$
                                             &$19.52^{+2.64}_{-3.47}$&$-2.79^{+1.20}_{-1.22}$&$18.95^{+2.87}_{-3.67} $\\ \hline

$B^0 \to (\rho^0 \to )\pi^+\pi^-(K^{*0}\to)K^+\pi^-$ &$23.55^{+3.59}_{-4.74}$&$-29.96^{+2.49}_{-2.44}$&$-3.20^{+1.55}_{-2.19}$
                                             &$26.76^{+2.62}_{-3.12}$&$-3.14^{+1.58}_{-2.04}$&$26.75^{+2.66}_{-3.20}$\\ \hline

$B^0 \to (\rho^- \to )\pi^-\pi^0(K^{*+}\to)K^0\pi^+$ &$19.50^{+2.41}_{-3.09}$&$-28.19^{+2.35}_{-2.02}$&$-4.34^{+1.02}_{-1.20}$
                                             &$23.84^{+1.94}_{-2.51}$&$-1.56^{+1.33}_{-1.70}$&$23.34^{+2.00}_{-2.73}$\\ \hline

$B^0_s\to (\rho^+ \to )\pi^+\pi^0(K^{*-}\to)\bar K^0\pi^-$ &$-3.79^{+1.61}_{-1.53}$&$3.72^{+0.76}_{-0.73}$&$-0.04^{+0.86}_{-0.88}$
                                              &$-3.75^{+0.91}_{-0.85}$&$0.30^{+0.85}_{-0.93}$&$-3.75^{+0.87}_{-0.81} $\\ \hline

$B^0_s\to (\rho^0 \to )\pi^+\pi^-(\bar{K}^{*0}\to)K^-\pi^+$ &$-30.96^{+1.31}_{-0.42}$&$29.56^{+2.57}_{-4.69}$&$-0.70^{+1.68}_{-2.10}$
                                                    &$-30.26^{+2.67}_{-0.94}$&$-8.97^{+3.06}_{-1.89}$&$-30.45^{+2.31}_{-0.51}$\\
\hline\hline
\multicolumn{1}{c|}{}  &\multicolumn{6}{|c}{$\text{TPAs}$-2}   \cr\cline{2-7}
{Modes} &$\mathcal{A}_{T}^2$&$\bar{\mathcal{A}}_{T}^2$ &$\mathcal{A}_{\text{T-true}}^2$ &$\mathcal{A}_{\text{T-fake}}^2$&$\mathcal{A}_{\text{T-True}}^{(2)\text{ave}}$ &$\mathcal{A}_{\text{T-fake}}^{(2)\text{ave}}$  \cr \hline
$B^+ \to (\rho^+ \to )\pi^+\pi^0(K^{*0}\to)K^+\pi^-$ &$-1.44^{+1.08}_{-1.07}$&$1.54^{+1.10}_{-1.04}$&$0.05^{+0.12}_{-0.04}$
                                             &$-1.49^{+1.06}_{-1.08}$&$0.04^{+0.11}_{-0.07}$&$-1.49^{+1.06}_{-1.08}$\\ \hline

$B^+ \to (\rho^0 \to )\pi^+\pi^-(K^{*+}\to)K^0\pi^+$ &$-1.18^{+1.32}_{-1.39}$&$-9.81^{+2.51}_{-2.56}$&$-5.49^{+1.63}_{-1.72}$
                                             &$4.31^{+1.14}_{-1.17}$&$-5.00^{+1.53}_{-1.61}$&$3.69^{+1.08}_{-1.16}$\\ \hline

$B^0 \to (\rho^0 \to )\pi^+\pi^-(K^{*0}\to)K^+\pi^-$ &$-1.73^{+1.51}_{-1.27}$&$16.27^{+3.73}_{-3.97}$&$7.27^{+2.07}_{-2.06}$
                                             &$-9.00^{+2.12}_{-1.93}$&$7.25^{+2.07}_{-2.18}$&$-8.98^{+2.21}_{-1.98}$\\ \hline

$B^0 \to (\rho^- \to )\pi^-\pi^0(K^{*+}\to)K^0\pi^+$ &$-0.93^{+0.65}_{-1.03}$&$1.55^{+1.05}_{-1.02}$&$0.31^{+0.25}_{-0.21}$
                                             &$-1.24^{+0.83}_{-0.83}$&$0.16^{+0.21}_{-0.11}$&$-1.20^{+0.80}_{-0.82}$\\ \hline

$B^0_s\to (\rho^+ \to )\pi^+\pi^0(K^{*-}\to)\bar K^0\pi^-$ &$-0.58^{+0.28}_{-0.28}$&$0.64^{+0.13}_{-0.16}$&$0.04^{+0.10}_{-0.16}$
                                              &$-0.61^{+0.18}_{-0.16}$&$0.08^{+0.12}_{-0.13}$&$-0.61^{+0.17}_{-0.16}$\\ \hline

$B^0_s\to (\rho^0 \to )\pi^+\pi^-(\bar{K}^{*0}\to)K^-\pi^+$ &$-3.43^{+0.49}_{-0.39}$&$1.80^{+0.45}_{-0.72}$&$-0.81^{+0.25}_{-0.37}$
                                                    &$-2.61^{+0.50}_{-0.35}$&$-1.53^{+0.39}_{-0.28}$&$-2.84^{+0.46}_{-0.29}$\\
\hline\hline
\end{tabular}
\end{threeparttable}
\end{center}
\end{table}
\section{CONCLUSION}\label{sec:4}

In this work, we have presented six helicity amplitudes of four-body decays $B_{(s)} \to (\pi\pi)(K\pi)$, where $\pi\pi$ invariant-mass spectrum is dominated by the vector $\rho$ resonance and scalar $f_0(500), f_0(980)$ resonances, and the vector $K^*$ resonance and scalar resonance $K_0^*(1430)$ are expected to contribute in the $K\pi$ invariant-mass range.
We have examined the branching ratios, polarization fractions, direct $CP$ asymmetries, triple product asymmetries in $B_{(s)} \to [VV,SS,SV,VS] \to (\pi\pi)(K\pi)$ decays.
In our numerical study, there exist many theoretical uncertainties in the calculation.
The uncertainties of the nonperturbative parameters of the two-meson DAs and  the variation of the hard scale provide the dominant theoretical errors  to the theoretical predictions for branching ratios
and other physical observables. Therefore, the relevant Gegenbauer moments should be further constrained to improve the precision of theoretical predictions and meet with future data.
In addition,  one should make a great effort to evaluate the  higher-order contributions to four-body $B$ meson decays in order to reduce the sensitivity to the variation of the hard scales.

We have extracted the branching ratios of two-body $B\rightarrow \rho K^*$  decays from the results for the
corresponding four-body decays under the narrow-width approximation and shown the polarization fractions of the related decay channels.
The obtained two-body branching ratios agree well with previous theoretical studies performed in the
two-body framework within errors.
The predicted hierarchy pattern for the longitudinal polarization fractions in the $B_{(s)}$ meson
decays into the $P$-wave $\pi\pi$ and $K\pi$ pairs is compatible with the data roughly.
However, there is a big gap between our prediction of longitudinal polarization fraction for $B^0 \to \rho^0 K^{*0}$ and the recent LHCb measurement, which should be resolved.
In addition, we have calculated the branching ratios of the four-body decays $B_{(s)} \to [SS,SV,VS]\to (\pi\pi)(K\pi)$.
For the decays associated with scalar resonance $f_0(500)$, we have used the BW and Bugg models
to parameterize the wide $f_0(500)$ meson respectively but found that the model-dependence of the PQCD predictions
is not significant.
The branching ratios of $B^0 \to (\rho^- \to )\pi^-\pi^0(K_0^{*}(1430)^{+}\to)K^0\pi^+$ and $B^0 \to (\rho^0 \to )\pi^+\pi^-(K_0^{*}(1430)^{0}\to)K^+\pi^-$  decays, which are related to isospin limit, remain puzzling and need to be resolved.

We have calculated the direct $CP$ asymmetries with each helicity state ($\mathcal{A}_{0,\parallel,\perp}^{\text{CP}}$) for the four-body $B_{(s)} \to VV \to (\pi\pi) (K\pi)$ decays, together with the direct $CP$ asymmetries of $B_{(s)} \to [SS,SV,VS] \to (\pi\pi)(K\pi)$ decays.
The $CP$ asymmetry in the four-body framework is dependent on the invariant mass of the final-state pairs, which results in the differences between the two-body and four-body frameworks in the PQCD approach.
Meanwhile, we perform an angular analysis on four-body $B_{(s)} \to \rho K^* \to (\pi\pi)(K\pi)$ decays to obtain the triple product asymmetries in detail.
We found that most ``true" TPAs are very small, which are consistent with the predictions of the standard model.
A ``true" TPA that is predicted to vanish provides an excellent place for looking for new physics because there is no suppression from the strong phases.

\begin{acknowledgments}
Many thanks to H.n.~Li for valuable discussions.
This work was supported by ``the Fundamental Research Funds for the Central Universities'' No.~KJQN202144 and the National Natural Science Foundation of China under the No.~12005103, No.~12075086, No.~11775117 and No.~11947013.
YL is also supported by the Natural Science Foundation of Jiangsu Province under Grant No.~BK20190508 and the Research Start-up Funding of Nanjing
Agricultural University.
DCY is also supported by the Natural Science Foundation of Jiangsu Province under Grant No.~BK20200980.
ZR is supported in part by the Natural Science Foundation of Hebei Province under Grant No.~A2019209449 and No.~A2021209002.

\end{acknowledgments}

\appendix
\section{Decay amplitudes}
In this Appendix we present the PQCD factorization formulas for the amplitudes of
the considered four-body hadronic $B$ meson decays:

\begin{itemize}
\item[]
$\bullet$ $ B \to \rho K^*\to(\pi\pi)(K\pi)$ decay modes ($h=0,\|,\perp$)
\begin{eqnarray}
A_h(B^0 \to (\rho^0 \to) \pi^+\pi^-(K^{*0}\to)K^+\pi^-)&=& \frac{G_F} {2}\big\{V_{ub}^*V_{us}[(C_1+\frac{C_2}{3})F^{LL,h}_{eK^*}+C_2M^{LL,h}_{eK^*}]\non
 &-&V_{tb}^*V_{ts}[\frac{3}{2}(C_7+\frac{C_8}{3}+C_9+\frac{C_{10}}{3})F^{LL,h}_{eK^*}\non
 &+&\frac{3C_{10}}{2}M^{LL,h}_{eK^*}+\frac{3C_8}{2}M^{SP,h}_{eK^*}\nonumber\\
 &-&(\frac{C_3}{3}+C_4-\frac{C_9}{6}-\frac{C_{10}}{2})(F^{LL,h}_{e\rho}+F^{LL,h}_{a\rho})\non
 &-&(\frac{C_5}{3}+C_6-\frac{C_7}{6}-\frac{C_8}{2})F^{SP,h}_{a\rho}\nonumber\\
&-&(C_3-\frac{C_9}{2})(M^{LL,h}_{e\rho}+M^{LL,h}_{a\rho})\non
&-&(C_5-\frac{C_7}{2})(M^{LR,h}_{e\rho}+M^{LR,h}_{a\rho})]\big\}  ,
\end{eqnarray}
\begin{eqnarray}
A_h(B^+ \to (\rho^+ \to) \pi^+\pi^0(K^{*0}\to)K^+\pi^-)&=& \frac{G_F} {\sqrt{2}}\big\{V_{ub}^*V_{us}[(\frac{C_1}{3}+C_2)F^{LL,h}_{a\rho}+C_1M^{LL,h}_{a\rho}]\non
&-&V_{tb}^*V_{ts}[(\frac{C_3}{3}+C_4-\frac{C_9}{6}-\frac{C_{10}}{2})F^{LL,h}_{e\rho}\non
&+&(C_3-\frac{C_9}{2})M^{LL,h}_{e\rho}+(C_5-\frac{C_7}{2})M^{LR,h}_{e\rho}\non
&+&(\frac{C_3}{3}+C_4+\frac{C_9}{3}+C_{10})F^{LL,h}_{a\rho}\non
&+&(\frac{C_5}{3}+C_6+\frac{C_7}{3}+C_{8})F^{SP,h}_{a\rho}\non
&+&(C_3+C_9)M^{LL,h}_{a\rho}+(C_5+C_7)M^{LR,h}_{a\rho}
]\big\},
\end{eqnarray}
\begin{eqnarray}
A_h(B^+ \to (\rho^0 \to) \pi^+\pi^-(K^{*+}\to)K^0\pi^+)&=& \frac{G_F} {2}\big\{V_{ub}^*V_{us}
[(\frac{C_1}{3}+C_2)(F^{LL,h}_{e\rho}+F^{LL,h}_{a\rho})+C_1(M^{LL,h}_{e\rho}+M^{LL,h}_{a\rho})\non
&+&C_2M^{LL,h}_{eK^*}+(C_1+\frac{C_2}{3})F^{LL,h}_{eK^*}]\non
&-&V_{tb}^*V_{ts}[(\frac{C_3}{3}+C_4+\frac{C_9}{3}+C_{10})(F^{LL,h}_{e\rho}+F^{LL,h}_{a\rho})\non
&+&(C_3+C_9)(M^{LL,h}_{e\rho}+M^{LL,h}_{a\rho})\non
&+&(C_5+C_7)(M^{LR,h}_{e\rho}+M^{LR,h}_{a\rho})\non
&+&(\frac{C_5}{3}+C_6+\frac{C_7}{3}+C_8)F^{SP,h}_{a\rho}\non
&+&\frac{3}{2}(C_7+\frac{C_8}{3}+C_9+\frac{C_{10}}{3})F^{LL,h}_{eK^*}\non
&+&\frac{3C_{10}}{2}M^{LL,h}_{eK^*}+\frac{3C_{8}}{2}M^{SP,h}_{eK^*}]\big\},
\end{eqnarray}
\begin{eqnarray}
A_h(B^0 \to (\rho^- \to) \pi^-\pi^0(K^{*+}\to)K^0\pi^+)&=& \frac{G_F} {\sqrt{2}}\big\{V_{ub}^*V_{us}[(\frac{C_1}{3}+C_2)F^{LL,h}_{e\rho}+C_1M^{LL,h}_{e\rho}]\non
&-&V_{tb}^*V_{ts}[(\frac{C_3}{3}+C_4+\frac{C_9}{3}+C_{10})F^{LL,h}_{e\rho}\non
&+&(C_3+C_9)M^{LL,h}_{e\rho}+(C_5+C_7)M^{LR,h}_{e\rho}\non
&+&(\frac{C_3}{3}+C_4-\frac{C_9}{6}-\frac{C_{10}}{2})F^{LL,h}_{a\rho}\non
&+&(\frac{C_5}{3}+C_6-\frac{C_7}{6}-\frac{C_{8}}{2})F^{SP,h}_{a\rho}\non
&+&(C_3-\frac{C_9}{2})M^{LL,h}_{a\rho}+(C_5-\frac{C_7}{2})M^{LR,h}_{a\rho}
]\big\},
\end{eqnarray}
\begin{eqnarray}
A_h(B_s^0 \to (\rho^+ \to) \pi^+\pi^0(K^{*-}\to)\bar K^0\pi^-)&=& \frac{G_F} {\sqrt{2}}\big\{V_{ub}^*V_{ud}[(\frac{C_1}{3}+C_2)F^{LL,h}_{eK^*}+C_1M^{LL,h}_{eK^*}]\non
&-&V_{tb}^*V_{td}[(\frac{C_3}{3}+C_4+\frac{C_9}{3}+C_{10})F^{LL,h}_{eK^*}\non
&+&(C_3+C_9)M^{LL,h}_{eK^*}+(C_5+C_7)M^{LR,h}_{eK^*}\non
&+&(\frac{C_3}{3}+C_4-\frac{C_9}{6}-\frac{C_{10}}{2})F^{LL,h}_{aK^*}\non
&+&(\frac{C_5}{3}+C_6-\frac{C_7}{6}-\frac{C_{8}}{2})F^{SP,h}_{aK^*}\non
&+&(C_3-\frac{C_9}{2})M^{LL,h}_{aK^*}+(C_5-\frac{C_7}{2})M^{LR,h}_{aK^*}
]\big\},
\end{eqnarray}
\begin{eqnarray}
A_h(B_s^0 \to (\rho^0 \to) \pi^+\pi^-(\bar{K}^{*0}\to)K^-\pi^+)&=& \frac{G_F} {2}\big\{V_{ub}^*V_{ud}[(C_1+\frac{C_2}{3})F^{LL,h}_{eK^*}+C_2M^{LL,h}_{eK^*}]\non
&-&V_{tb}^*V_{td}[(-\frac{C_3}{3}-C_4+\frac{3C_7}{2}+\frac{C_8}{2}+\frac{5C_9}{3}+C_{10})F^{LL,h}_{eK^*}\non
&+&(-C_3+\frac{C_9}{2}+\frac{3C_{10}}{2})M^{LL,h}_{eK^*}\non
&-&(C_5-\frac{C_7}{2})M^{LR,h}_{eK^*}+\frac{3C_{8}}{2}M^{SP,h}_{eK^*}\non
&-&(\frac{C_3}{3}+C_4-\frac{C_9}{6}-\frac{C_{10}}{2})F^{LL,h}_{aK^*}\non
&-&(\frac{C_5}{3}+C_6-\frac{C_7}{6}-\frac{C_{8}}{2})F^{SP,h}_{aK^*}\non
&-&(C_3-\frac{C_9}{2})M^{LL,h}_{aK^*}-(C_5-\frac{C_7}{2})M^{LR,h}_{aK^*}
]\big\},
\end{eqnarray}
\item[]
$\bullet$ $ B \to f_0 K^*\to(\pi\pi)(K\pi)$ decay modes
\begin{eqnarray}
A(B^+ \to (f_0(980) \to) \pi^+\pi^- (K^{*+}\to)K^0\pi^+)&=& \frac{G_F}{\sqrt{2}}\big\{V_{ub}^*V_{us}
[(\frac{C_1}{3}+C_2)F^{LL}_{aK^*}+C_1M^{LL}_{aK^*}]\non
&-&V_{tb}^*V_{ts}[(\frac{C_5}{3}+C_6-\frac{C_7}{6}-\frac{C_8}{2})F^{SP}_{eK^*}\non
&+&(C_3+C_4-\frac{1}{2}(C_9+C_{10}))M^{LL}_{eK^*}+(C_5-\frac{C_7}{2})M^{LR}_{eK^*}\non
&+&(C_6-\frac{C_8}{2})M^{SP}_{eK^*}+(\frac{C_3}{3}+C_4+\frac{C_9}{3}+C_{10})F^{LL}_{aK^*}\non
&+&(\frac{C_5}{3}+C_6+\frac{C_7}{3}+C_{8})F^{SP}_{aK^*}+(C_3+C_9)M^{LL}_{eK^*}\non
&+& (C_5+C_7)M^{LR}_{eK^*}]\big\},
\end{eqnarray}
\begin{eqnarray}
A(B^0 \to (f_0(500) \to) \pi^+\pi^- (K^{*0}\to)K^+\pi^-)&=& \frac{G_F}{2}\big\{V_{ub}^*V_{us}
[C_2M^{LL}_{eK^*}]\non
&-&V_{tb}^*V_{ts}[(\frac{C_3}{3}+C_4-\frac{C_9}{6}-\frac{C_{10}}{2})(F^{LL}_{ef_0}+F^{LL}_{af_0})\non
&+&(C_3-\frac{C_9}{2})(M^{LL}_{ef_0}+M^{LL}_{af_0})+(C_5-\frac{C_7}{2})(M^{LR}_{ef_0}+M^{LR}_{af_0})\non
&+&(\frac{C_5}{3}+C_6-\frac{C_7}{6}-\frac{C_{8}}{2})F^{SP}_{af_0}\non
&+&(2C_4+\frac{C_{10}}{2})M^{LL}_{eK^*}+(2C_6+\frac{C_{8}}{2})M^{SP}_{eK^*}]\big\},
\end{eqnarray}
\begin{eqnarray}
A(B_s^0 \to (f_0(500) \to) \pi^+\pi^-(\bar K^{*0}\to)K^-\pi^+)&=&\frac{G_F}{2}\big\{V_{ub}^*V_{ud}[C_2M^{LL}_{eK^*}]\non
 &-&V_{tb}^*V_{td}[(\frac{C_5}{3}+C_6-\frac{C_7}{6}-\frac{C_8}{2})(F^{SP}_{eK^*}+F^{SP}_{aK^*})\non
&+&(C_3+2C_4-\frac{1}{2}(C_9-C_{10}))M^{LL}_{eK^*}+(C_5-\frac{C_7}{2})M^{LR}_{eK^*}\non
&+&(2C_6+\frac{C_8}{2})M^{SP}_{eK^*}+(\frac{C_3}{3}+C_4-\frac{C_9}{6}-\frac{C_{10}}{2})F^{LL}_{aK^*}\non
&+&(C_3-\frac{C_9}{2})M^{LL}_{eK^*}+ (C_5-\frac{C_7}{2})M^{LR}_{eK^*}]\big\},
\end{eqnarray}
\begin{eqnarray}
A(B^0 \to (f_0(980) \to) \pi^+\pi^- (K^{*0}\to)K^+\pi^-)&=& -\frac{G_F}{\sqrt{2}}
V_{tb}^*V_{ts}[(\frac{C_5}{3}+C_6-\frac{C_7}{6}-\frac{C_8}{2})(F^{SP}_{eK^*}+F^{SP}_{aK^*})\non
&+&(C_3+C_4-\frac{1}{2}(C_9+C_{10}))M^{LL}_{eK^*}+(C_5-\frac{C_7}{2})M^{LR}_{eK^*}\non
&+&(C_6-\frac{C_8}{2})M^{SP}_{eK^*}+(\frac{C_3}{3}+C_4-\frac{C_9}{6}-\frac{C_{10}}{2})F^{LL}_{aK^*}\non
&+&(C_3-\frac{C_9}{2})M^{LL}_{eK^*}+ (C_5-\frac{C_7}{2})M^{LR}_{eK^*}]\big\},
\end{eqnarray}
\begin{eqnarray}
A(B_s^0 \to (f_0(980) \to) \pi^+\pi^-(\bar K^{*0}\to)K^-\pi^+)&=& -\frac{G_F}{\sqrt{2}}
V_{tb}^*V_{td}[(\frac{C_3}{3}+C_4-\frac{C_9}{6}-\frac{C_{10}}{2})(F^{LL}_{ef_0}+F^{LL}_{af_0})\non
&+&(C_3-\frac{C_9}{2})(M^{LL}_{ef_0}+M^{LL}_{af_0})+(C_5-\frac{C_7}{2})(M^{LR}_{ef_0}+M^{LR}_{af_0})\non
&+&(\frac{C_5}{3}+C_6-\frac{C_7}{6}-\frac{C_8}{2})F^{SP}_{af_0}+(C_4-\frac{C_{10}}{2})M^{LL}_{eK^*}\non
&+&(C_6-\frac{C_8}{2})M^{SP}_{eK^*}]\big\},
\end{eqnarray}
\begin{eqnarray}\label{A12}
A(B^+ \to (f_0(500) \to) \pi^+\pi^-(K^{*+}\to)K^0\pi^+)&=& \frac{G_F}{2}\big\{V_{ub}^*V_{us}
[(C_2+\frac{C_1}{3})(F^{LL}_{ef_0}+F^{LL}_{af_0})+C_1(M^{LL}_{ef_0}+M^{LL}_{af_0})\non
&+&C_2M^{LL}_{eK^*}]-V_{tb}^*V_{ts}[(\frac{C_3}{3}+C_4+\frac{C_9}{3}+C_{10})(F^{LL}_{ef_0}+F^{LL}_{af_0})\non
&+&(C_3+C_9)(M^{LL}_{ef_0}+M^{LL}_{af_0})+(C_5+C_7)(M^{LR}_{ef_0}+M^{LR}_{af_0})\non
&+&(\frac{C_5}{3}+C_6+\frac{C_7}{3}+C_8)F^{SP}_{af_0}+(2C_4+\frac{C_{10}}{2})M^{LL}_{eK^*}\non
&+&(2C_6+\frac{C_{8}}{2})M^{SP}_{eK^*}]\big\},
\end{eqnarray}
\item[]
$\bullet$ $B \to f_0 K_0^*(1430)\to(\pi\pi)(K\pi)$ decay modes
\begin{eqnarray}
A(B^+ \to (f_0(500) \to) \pi^+\pi^- (K_0^*(1430)^+\to)K^0\pi^+)&=& \frac{G_F}{2}\big\{V_{ub}^*V_{us}[
(\frac{C_1}{3}+C_2)(F^{LL}_{ef_0}+F^{LL}_{af_0})\non
&+&C_1(M^{LL}_{ef_0}+M^{LL}_{af_0})+C_2M^{LL}_{eK_0^*}]\non
&-&V_{tb}^*V_{ts}[(\frac{C_3}{3}+C_4+\frac{C_9}{3}+C_{10})(F^{LL}_{ef_0}+F^{LL}_{af_0})\non
&+&(\frac{C_5}{3}+C_6+\frac{C_7}{3}+C_{8})(F^{SP}_{ef_0}+F^{SP}_{af_0})\non
&+&(C_3+C_9)(M^{LL}_{ef_0}+M^{LL}_{af_0})+(C_5+C_7)(M^{LR}_{ef_0}+M^{LR}_{af_0})\non
&+&(2C_4+\frac{C_{10}}{2})M^{LL}_{eK_0^*}+(2C_6+\frac{C_{8}}{2})M^{SP}_{eK_0^*}]\big\},
\end{eqnarray}
\begin{eqnarray}
A(B^0 \to (f_0(500) \to) \pi^+\pi^-( K^*_0(1430)^0\to)K^+\pi^-)&=& \frac{G_F}{2}\big\{V_{ub}^*V_{us}[C_2M^{LL}_{eK_0^*}]\non
&-&V_{tb}^*V_{ts}[(\frac{C_3}{3}+C_4-\frac{C_9}{6}-\frac{C_{10}}{2})(F^{LL}_{ef_0}+F^{LL}_{af_0})\non
&+&(\frac{C_5}{3}+C_6-\frac{C_7}{6}-\frac{C_{8}}{2})(F^{SP}_{ef_0}+F^{SP}_{af_0})\non
&+&(C_3-\frac{C_9}{2})(M^{LL}_{ef_0}+M^{LL}_{af_0})+(C_5-\frac{C_7}{2})(M^{LR}_{ef_0}+M^{LR}_{af_0})\non
&+&(2C_4+\frac{C_{10}}{2})M^{LL}_{eK_0^*}+(2C_6+\frac{C_{8}}{2})M^{SP}_{eK_0^*}]\big\},
\end{eqnarray}
\begin{eqnarray}
A(B_s^0 \to (f_0(500) \to )\pi^+\pi^-(\bar K_0^*(1430)^0\to)K^-\pi^+)&=&\frac{G_F}{2}\big\{V_{ub}^*V_{ud}[C_2M^{LL}_{eK_0^*}]\non
 &-&V_{tb}^*V_{td}[(\frac{C_5}{3}+C_6-\frac{C_7}{6}-\frac{C_8}{2})(F^{SP}_{eK^*_0}+F^{SP}_{aK^*_0})\non
&+&(C_3+2C_4-\frac{1}{2}(C_9-C_{10}))M^{LL}_{eK^*_0}+(C_5-\frac{C_7}{2})M^{LR}_{eK^*_0}\non
&+&(2C_6+\frac{C_8}{2})M^{SP}_{eK^*_0}+(\frac{C_3}{3}+C_4-\frac{C_9}{6}-\frac{C_{10}}{2})F^{LL}_{aK^*_0}\non
&+&(C_3-\frac{C_9}{2})M^{LL}_{eK^*_0}+ (C_5-\frac{C_7}{2})M^{LR}_{eK^*_0}]\big\},
\end{eqnarray}
\begin{eqnarray}
A(B^+ \to (f_0(980) \to)\pi^+\pi^-(K^{*}_0(1430)^+\to)K^0\pi^+)&=& \frac{G_F}{\sqrt{2}}\big\{V_{ub}^*V_{us}
[(\frac{C_1}{3}+C_2)F^{LL}_{aK^{*}_0}+C_1M^{LL}_{aK^{*}_0}]\non
&-&V_{tb}^*V_{ts}[(\frac{C_5}{3}+C_6-\frac{C_7}{6}-\frac{C_8}{2})F^{SP}_{eK^{*}_0}\non
&+&(C_3+C_4-\frac{1}{2}(C_9+C_{10}))M^{LL}_{eK^{*}_0}+(C_5-\frac{C_7}{2})M^{LR}_{eK^{*}_0}\non
&+&(C_6-\frac{C_8}{2})M^{SP}_{eK^{*}_0}+(\frac{C_3}{3}+C_4+\frac{C_9}{3}+C_{10})F^{LL}_{aK^{*}_0}\non
&+&(\frac{C_5}{3}+C_6+\frac{C_7}{3}+C_{8})F^{SP}_{aK^{*}_0}+(C_3+C_9)M^{LL}_{eK^{*}_0}\non
&+& (C_5+C_7)M^{LR}_{eK^{*}_0}]\big\},
\end{eqnarray}
\begin{eqnarray}
A(B^0 \to (f_0(980) \to) \pi^+\pi^-(K_0^*(1430)^0\to)K^+\pi^-)&=& -\frac{G_F}{\sqrt{2}}
V_{tb}^*V_{ts}[(\frac{C_5}{3}+C_6-\frac{C_7}{6}-\frac{C_8}{2})(F^{SP}_{eK^*_0}+F^{SP}_{aK^*_0})\non
&+&(C_3+C_4-\frac{1}{2}(C_9+C_{10}))M^{LL}_{eK^*_0}+(C_5-\frac{C_7}{2})M^{LR}_{eK^*_0}\non
&+&(C_6-\frac{C_8}{2})M^{SP}_{eK^*_0}+(\frac{C_3}{3}+C_4-\frac{C_9}{6}-\frac{C_{10}}{2})F^{LL}_{aK^*_0}\non
&+&(C_3-\frac{C_9}{2})M^{LL}_{eK^*_0}+ (C_5-\frac{C_7}{2})M^{LR}_{eK^*_0}]\big\},
\end{eqnarray}
\begin{eqnarray}
A(B_s^0 \to (f_0(980) \to )\pi^+\pi^- (\bar K_0^*(1430)^0\to)K^-\pi^+)&=& -\frac{G_F}{\sqrt{2}}
V_{tb}^*V_{td}[(\frac{C_3}{3}+C_4-\frac{C_9}{6}-\frac{C_{10}}{2})(F^{LL}_{ef_0}+F^{LL}_{af_0})\non
&+&(C_3-\frac{C_9}{2})(M^{LL}_{ef_0}+M^{LL}_{af_0})+(C_5-\frac{C_7}{2})(M^{LR}_{ef_0}+M^{LR}_{af_0})\non
&+&(\frac{C_5}{3}+C_6-\frac{C_7}{6}-\frac{C_8}{2})(F^{SP}_{ef_0}+F^{SP}_{af_0})+(C_4-\frac{C_{10}}{2})M^{LL}_{eK^*_0}\non
&+&(C_6-\frac{C_8}{2})M^{SP}_{eK^*_0}]\big\},
\end{eqnarray}
\item[]
$\bullet$ $B \to \rho K_0^*(1430)\to(\pi\pi)(K\pi)$ decay modes
\begin{eqnarray}
A(B^0 \to (\rho^0 \to )\pi^+\pi^- (K_0^*(1430)^0\to)K^+\pi^-)&=& \frac{G_F} {2}\big\{V_{ub}^*V_{us}[(C_1+\frac{C_2}{3})F^{LL}_{eK_0^*}+C_2M^{LL}_{eK_0^*}]\non
 &-&V_{tb}^*V_{ts}[\frac{3}{2}(C_7+\frac{C_8}{3}+C_9+\frac{C_{10}}{3})F^{LL}_{eK_0^*}\non
 &+&\frac{3C_{10}}{2}M^{LL}_{eK_0^*}+\frac{3C_8}{2}M^{SP}_{eK_0^*}\nonumber\\
 &-&(\frac{C_3}{3}+C_4-\frac{C_9}{6}-\frac{C_{10}}{2})(F^{LL}_{e\rho}+F^{LL}_{a\rho})\non
 &-&(C_3-\frac{C_9}{2})(M^{LL}_{e\rho}+M^{LL}_{a\rho})\non
 &-&(\frac{C_5}{3}+C_6-\frac{C_7}{6}-\frac{C_8}{2})(F^{SP}_{e\rho}+F^{SP}_{a\rho})\non
 &-&(C_5-\frac{C_7}{2})(M^{LR}_{e\rho}+M^{LR}_{a\rho})]\big\}, \non \label{amp13}
\end{eqnarray}
\begin{eqnarray}
A(B^+ \to (\rho^+ \to )\pi^+\pi^0 (K_0^*(1430)^0\to)K^+\pi^-)&=& \frac{G_F} {\sqrt{2}}\big\{V_{ub}^*V_{us}[(\frac{C_1}{3}+C_2)F^{LL}_{a\rho}+C_1M^{LL}_{a\rho}]\non
&-&V_{tb}^*V_{ts}[(\frac{C_3}{3}+C_4-\frac{C_9}{6}-\frac{C_{10}}{2})F^{LL}_{e\rho}\non
&+&(\frac{C_5}{3}+C_6-\frac{C_7}{6}-\frac{C_8}{2})F^{SP}_{e\rho}\non
&+&(C_3-\frac{C_9}{2})M^{LL}_{e\rho}+(C_5-\frac{C_7}{2})M^{LR}_{e\rho}\non
&+&(\frac{C_3}{3}+C_4+\frac{C_9}{3}+C_{10})F^{LL}_{a\rho}\non
&+&(\frac{C_5}{3}+C_6+\frac{C_7}{3}+C_{8})F^{SP}_{a\rho}\non
&+&(C_3+C_9)M^{LL}_{a\rho}+(C_5+C_7)M^{LR}_{a\rho}
]\big\},
\end{eqnarray}
\begin{eqnarray}
A(B^+ \to (\rho^0 \to) \pi^+\pi^- (K_0^*(1430)^+\to)K^0\pi^+)&=& \frac{G_F} {2}\big\{V_{ub}^*V_{us}
[(\frac{C_1}{3}+C_2)(F^{LL}_{e\rho}+F^{LL}_{a\rho})\non
&+&C_1(M^{LL}_{e\rho}+M^{LL}_{a\rho})+C_2M^{LL}_{eK_0^*}+(C_1+\frac{C_2}{3})F^{LL}_{eK_0^*}]\non
&-&V_{tb}^*V_{ts}[(\frac{C_3}{3}+C_4+\frac{C_9}{3}+C_{10})(F^{LL}_{e\rho}+F^{LL}_{a\rho})\non
&+&(\frac{C_5}{3}+C_6+\frac{C_7}{3}+C_8)(F^{SP}_{e\rho}+F^{SP}_{a\rho})\non
&+&(C_3+C_9)(M^{LL}_{e\rho}+M^{LL}_{a\rho})\non
&+&(C_5+C_7)(M^{LR}_{e\rho}+M^{LR}_{a\rho})\non
&+&\frac{3}{2}(C_7+\frac{C_8}{3}+C_9+\frac{C_{10}}{3})F^{LL}_{eK_0^*}\non
&+&\frac{3C_{10}}{2}M^{LL}_{eK_0^*}+\frac{3C_{8}}{2}M^{SP}_{eK_0^*}]\big\},
\end{eqnarray}
\begin{eqnarray}
A(B^0 \to (\rho^- \to )\pi^-\pi^0 (K_0^*(1430)^+\to)K^0\pi^+)&=& \frac{G_F} {\sqrt{2}}\big\{V_{ub}^*V_{us}[(\frac{C_1}{3}+C_2)F^{LL}_{e\rho}+C_1M^{LL}_{e\rho}]\non
&-&V_{tb}^*V_{ts}[(\frac{C_3}{3}+C_4+\frac{C_9}{3}+C_{10})F^{LL}_{e\rho}\non
&+&(\frac{C_5}{3}+C_6+\frac{C_7}{3}+C_8)F^{SP}_{e\rho}\non
&+&(C_3+C_9)M^{LL}_{e\rho}+(C_5+C_7)M^{LR}_{e\rho}\non
&+&(\frac{C_3}{3}+C_4-\frac{C_9}{6}-\frac{C_{10}}{2})F^{LL}_{a\rho}\non
&+&(\frac{C_5}{3}+C_6-\frac{C_7}{6}-\frac{C_{8}}{2})F^{SP}_{a\rho}\non
&+&(C_3-\frac{C_9}{2})M^{LL}_{a\rho}+(C_5-\frac{C_7}{2})M^{LR}_{a\rho}
]\big\},
\end{eqnarray}
\begin{eqnarray}
A(B_s^0 \to (\rho^+ \to) \pi^+\pi^0 (K_0^*(1430)^-\to)\bar K^0\pi^-)&=& \frac{G_F} {\sqrt{2}}\big\{V_{ub}^*V_{ud}[(\frac{C_1}{3}+C_2)F^{LL}_{eK_0^*}+C_1M^{LL}_{eK_0^*}]\non
&-&V_{tb}^*V_{td}[(\frac{C_3}{3}+C_4+\frac{C_9}{3}+C_{10})F^{LL}_{eK_0^*}\non
&+&(\frac{C_5}{3}+C_6+\frac{C_7}{3}+C_8)F^{SP}_{eK_0^*}\non
&+&(C_3+C_9)M^{LL}_{eK_0^*}+(C_5+C_7)M^{LR}_{eK_0^*}\non
&+&(\frac{C_3}{3}+C_4-\frac{C_9}{6}-\frac{C_{10}}{2})F^{LL}_{aK_0^*}\non
&+&(\frac{C_5}{3}+C_6-\frac{C_7}{6}-\frac{C_{8}}{2})F^{SP}_{aK_0^*}\non
&+&(C_3-\frac{C_9}{2})M^{LL}_{aK_0^*}+(C_5-\frac{C_7}{2})M^{LR}_{aK_0^*}]\big\},
\end{eqnarray}
\begin{eqnarray}
A(B_s^0 \to (\rho^0 \to )\pi^+\pi^- (\bar{K}_0^*(1430)^0\to)K^-\pi^+)&=& \frac{G_F} {2}\big\{V_{ub}^*V_{ud}[(C_1+\frac{C_2}{3})F^{LL}_{eK_0^*}+C_2M^{LL}_{eK_0^*}]\non
&-&V_{tb}^*V_{td}[(-\frac{C_3}{3}-C_4+\frac{3C_7}{2}+\frac{C_8}{2}+\frac{5C_9}{3}+C_{10})F^{LL}_{eK_0^*}\non
&-&(\frac{C_5}{3}+C_6-\frac{C_7}{6}-\frac{C_{8}}{2})F^{SP}_{eK_0^*}\non
&+&(-C_3+\frac{C_9}{2}+\frac{3C_{10}}{2})M^{LL}_{eK_0^*}\non
&-&(C_5-\frac{C_7}{2})M^{LR}_{eK^*}+\frac{3C_{8}}{2}M^{SP}_{eK_0^*}\non
&-&(\frac{C_3}{3}+C_4-\frac{C_9}{6}-\frac{C_{10}}{2})F^{LL}_{aK_0^*}\non
&-&(\frac{C_5}{3}+C_6-\frac{C_7}{6}-\frac{C_{8}}{2})F^{SP}_{aK_0^*}\non
&-&(C_3-\frac{C_9}{2})M^{LL}_{aK_0^*}-(C_5-\frac{C_7}{2})M^{LR}_{aK_0^*}]\big\},
\end{eqnarray}
\end{itemize}
where $G_F=1.16639\times 10^{-5}$ GeV$^{-2}$ is the Fermi coupling constant and the $V_{ij}$'s are the Cabibbo-Kobayashi-Maskawa matrix elements.
The superscripts $LL$, $LR$, and $SP$ refer to the contributions from $(V-A)\otimes(V-A)$, $(V-A)\otimes(V+A)$, and $(S-P)\otimes(S+P)$ operators, respectively.
The explicit formulas for the factorizable emission (annihilation) contributions $F_{e(a)}$ and the nonfactorizable emission (annihilation) contributions
$M_{e(a)}$ from Fig.~\ref{fig2} can be obtained easily in Ref.~\cite{Rui:2021kbn}.

\end{document}